\documentclass[a4paper,10pt,notitlepage,twocolumn]{revtex4-1}

\usepackage{graphicx,color}
\usepackage[colorlinks=true,linkcolor=blue,citecolor=blue]{hyperref}
\usepackage{amssymb,amsmath,bbold,mathtools}

\newcommand{\rev}[1]{\textcolor{black}{#1}}
\newcommand{\revtwo}[1]{\textcolor{black}{#1}}

\begin{document}

\title{\rev{Emergent bosons} in the fermionic two-leg flux ladder}
\author{Marcello Calvanese Strinati}
\author{Richard Berkovits}
\author{Efrat Shimshoni}
\affiliation{Department of Physics, Bar-Ilan University, 52900 Ramat-Gan, Israel}
\date{\today}

\begin{abstract}
We study the \rev{emergence of bosonic pairs} in a system of two coupled one-dimensional \rev{fermionic} chains subject to a gauge flux (two-leg flux ladder), with both attractive and repulsive interaction. In the presence of strong attractive nearest-neighbour interaction and repulsive next-to-nearest-neighbour interaction, the system crosses into a regime in which fermions form tightly-bound pairs, which behave as bosonic entities. By means of numerical simulations based on the density-matrix-renormalization-group (DMRG) method, we show in particular that in the strongly-paired regime, the gauge flux induces a quantum phase transition of the Ising type from vortex density wave (VDW) to a charge density wave (CDW), characteristic of bosonic systems.
\end{abstract}

\maketitle

\section{Introduction}
Exotic phases of matter emerging from the interplay between strong interactions, magnetic fields and enhanced quantum fluctuations due to low dimensionality have been an active field of research in condensed-matter physics during the last decades, both for fermionic and bosonic systems. In the last years, a renewed theoretical and experimental interest in the realization and characterization of such intriguing phases has been triggered by the advances in the field of ultra-cold atomic gases in optical lattices with artificial gauge fields, the latter mimicking the effects of applied magnetic fields~\cite{bloch2005ultracoldatoms,doi:10.1080/00018730701223200,RevModPhys.80.885,RevModPhys.83.1523,Boada_2015,goldman2016topological,PhysRevLett.121.150403}. Such techniques provide the ability of creating and manipulating matter (\emph{synthetic matter}) with unprecedented precision.

In this respect, systems of many coupled one-dimensional (1D) chains immersed in a gauge field (\emph{flux ladders}) represent a versatile platform in which such effects can be studied, in which dimensionality is controlled by the number of wires. Because of their 1D nature, the toolbox to theoretically analyze  phases in these systems is provided by \emph{ad-hoc} numerical algorithms based on the density-matrix-renormalization-group (DMRG)~\cite{PhysRevLett.69.2863,RevModPhys.77.259} or matrix-product-state (MPS)~\cite{SCHOLLWOCK201196} formalism, and effective field theories, such as bosonization~\cite{gogolin2004bosonization,giamarchi2003quantum}.

The minimal setup in which gauge-field effects can be obtained is the two-leg flux ladder, i.e., two connected chains. Several works have focused on this system, discussing interesting aspects both for bosons~\cite{PhysRevB.64.144515,PhysRevA.85.041602,PhysRevB.84.054517,PhysRevB.83.220518,petrescu2013bosonic,PhysRevB.87.174501,PhysRevA.89.063617,1367-2630-16-7-073005,DiDio2015,PhysRevB.91.140406,PhysRevB.92.060506,1367-2630-17-9-092001,PhysRevA.92.053623,PhysRevLett.115.190402,PhysRevA.94.063628,1367-2630-18-5-055017,Strinati_2018,PhysRevA.97.033619,PhysRevA.98.033605,PhysRevA.99.053601} and fermions~\cite{PhysRevB.71.161101,PhysRevB.73.195114,1367-2630-17-10-105001,ncomms9134,1367-2630-18-3-035010,PhysRevA.95.063612,PhysRevLett.118.230402,PhysRevA.93.013604,PhysRevA.93.023608,Haller_2018}. In particular, it has been shown that flux ladders can host phases that, at low energies, are analogous to fractional quantum Hall (FQH) phases~\cite{petrescu2015chiral,cornfeld2015chiral,strinati2017laughlin,petrescu2017precursor,PhysRevB.99.245101}, or manifest quantum phase transitions from superconducting (SC) to Mott insulating phases~\cite{PhysRevB.64.144515,PhysRevB.83.220518,PhysRevB.91.140406,1367-2630-18-5-055017,DiDio2015}.

\begin{figure}[b]
\centering
\includegraphics[width=8cm]{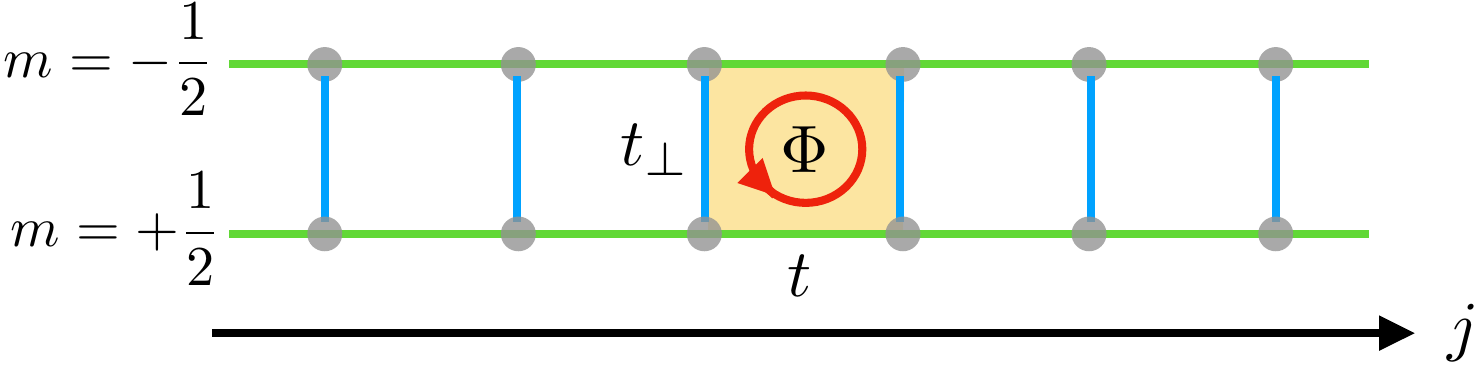}
\caption{Scheme of the two-leg flux ladder. The system consists of two 1D chains, labelled by $m=\pm1/2$, of $L$ sites each (grey dots), labeled by $j=1,\ldots,L$; $t$ (green bonds) and $t_\perp$ (blue bonds) are the intra- and inter-leg hopping parameters, respectively; $\Phi$ is the gauge flux per \emph{plaquette} (yellow area). \rev{A pair of sites with equal $j$ identifies a \emph{rung}.}}
\label{fig:schemeoftheladder}
\end{figure}

The phase diagram of the fermionic and bosonic two-leg flux ladder has been discussed in details for different models of interactions~\cite{PhysRevB.73.195114,CARR201322,PhysRevA.92.053623}, both attractive and repulsive. While it is expected that attractive on-site interactions in the fermionic ladder lead to the formation of fermionic pairs, which behave as bosonic particles~\cite{PhysRevB.64.144515,giamarchi2003quantum}, to the best of our knowledge, a detailed study of how such bosons emerge in the fermionic ladder for longer-range interactions is still missing. In this \revtwo{paper}, we aim to bridge this gap, studying the emergence of bosonic particles in the fermionic two-leg flux ladder with attractive and repulsive finite-range interactions.

\revtwo{This paper is organized as follows. We introduce our microscopic model in Sec.~\ref{sec:model}, and discuss its low-energy theory in the strongly-interacting regime in Sec.~\ref{sec:lownenergytheoriinteractions}. We then present our numerical results in Sec.~\ref{sec:numericalresults}. We draw our conclusions in Sec.~\ref{sec:conclusions}, and present additional numerical data in the Appendixes.}

\section{Model}
\label{sec:model}
The system consists of two 1D chains immersed in a gauge flux (Fig.~\ref{fig:schemeoftheladder}). In the following, we consider open boundary conditions (OBC) both in the longitudinal ($j$) and transverse ($m$) dimensions, and model our system by the Hamiltonian $\hat H_F=\hat H_0+\hat H_\perp+\hat H_{\rm int}$, where
\begin{equation}
\hat H_0\!=\!-t\,\sum_{j=1}^{L-1}\,\sum_{m=\pm1/2}\,\hat c^\dag_{j,m}\hat c_{j+1,m}+{\rm H.c.}
\end{equation}
\vspace{-0.5cm}
\begin{equation}
\hat H_\perp\!=\!t_\perp\,\sum_{j=1}^{L}e^{-i\Phi j}\,\hat c^\dag_{j,-\frac{1}{2}}\hat c_{j,+\frac{1}{2}}+{\rm H.c.}
\end{equation}
\vspace{-0.5cm}
\begin{equation}
\hat H_{\rm int}\!=\!\!\sum_{m=\pm1/2}\!\!\left(\!V\sum_{j=1}^{L-1}\hat n_{j,m}\hat n_{j+1,m}+W\sum_{j=1}^{L-2}\hat n_{j,m}\hat n_{j+2,m}\!\right)
\label{eq:interactionhamiltonian}
\end{equation}
in which we set the lattice constant to unity. In the Hamiltonian, $\hat c_{j,m}$ ($\hat c^\dag_{j,m}$) is the annihilation (creation) operator of a fermion on site $j$ and on leg $m$, and $\hat n_{j,m}=\hat c^\dag_{j,m}\hat c_{j,m}$ is the fermionic density operator; $t$ and $t_\perp$ denote the longitudinal and transverse hopping parameters, respectively, and $\Phi$ is the gauge flux per plaquette. We denote by $L$ the number of \rev{rungs of the ladder}, and $N$ the total number of particles in the system. We define the total particle density as $n=N/L$. The interaction Hamiltonian $\hat H_{\rm int}$ accounts for both intra-leg nearest-neighbour (NN) and next-to-nearest-neighbour (NNN) interaction, whose strengths are identified by $V$ and $W$, respectively. In the following, if not explicit, we use $t$ as reference energy scale. Since we aim at forming fermionic pairs, we consider $V<0$ and $W>0$. The first condition induces NN particles to bind, whereas the second one prevents clusters from forming.

\section{Low-energy theory for strong interactions}
\label{sec:lownenergytheoriinteractions}
\revtwo{In this section, we derive the low-energy theory of the model introduced in Sec.~\ref{sec:model}, in the strongly-interacting regime.} For sufficiently large but finite $|V|$ and $W$, which is the case of interest, the resulting ground state (GS) is composed of tightly-bound fermionic pairs \rev{subject to a hard pairing gap~\cite{PhysRevB.96.085133,borla2019confinedphases}} \revtwo{(see also Appendix~\ref{sec:directevidencepairinggap})}, with effective hopping parameters $\tilde t\sim t^2/|V|$ and $\tilde t_\perp\sim t_\perp^2/|V|$. \rev{In this strongly-coupled limit, we} bosonize the model starting from the fermionic pair operator $\hat B^\dag_{j,m}=\hat c^\dag_{j,m}\hat c^\dag_{j+1,m}$.

\begin{figure}[t]
\centering
\includegraphics[width=8cm]{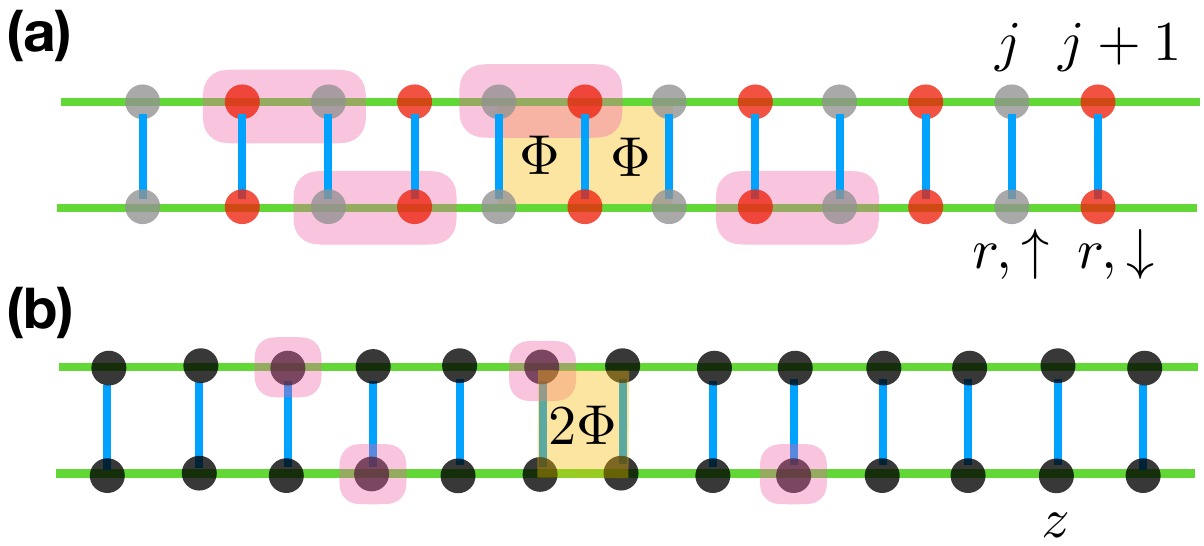}
\caption{Pictorial representation of the remapping of the two-leg ladder in Fig.~\ref{fig:schemeoftheladder}. \textbf{(a)} The chain is divided into two sublattices (grey and red dots) comprising of even and odd sites, identified by $\alpha=\uparrow,\downarrow$. Each pair of sites $j,j+1$ is split into a new set of variables $(r,\uparrow),(r,\downarrow)$. A pair (purple area) can be localized either at $(r,\uparrow),(r,\downarrow)$, or at $(r,\downarrow),(r+1,\uparrow)$. \textbf{(b)} A pair can be coarse-grained into a one-site bosonic particle localized at the center of mass $z=j+1/2$ of the fermionic pair (black dots), with a flux per plaquette equal to $2\Phi$.}
\label{fig:schemeoftheladderremapped}
\end{figure}

Since the NN and NNN interactions couple site with different and equal parity, respectively, one can interpret each chain of length $L$ (suppose $L$ even) as the composition of two sublattices of length $L/2$ each, identified by a pseudo-spin index $\alpha=\uparrow,\downarrow$ and lattice coordinate $r$, such that $\alpha=\uparrow$ includes the sites of the of the original lattice with $j$ odd, and $\alpha=\downarrow$ includes those with $j$ even [Fig.~\ref{fig:schemeoftheladderremapped}(a)]: $j=2r-R(\alpha)$, where $R(\uparrow)=1$ and $R(\downarrow)=0$. The fermionic lattice operator is then recast as $\hat c_{j,m}\rightarrow\hat c_{r,\alpha,m}$, whose bosonized version reads~\cite{giamarchi2003quantum,RevModPhys.83.1405}
\begin{equation}
\hat c_{r,\alpha,m}\sim\sum_{p}e^{-i\sqrt{\pi}\,\hat\theta_{\alpha,m}(r)}\,e^{-ip\pi(n/2)r}\,e^{ip\sqrt{\pi}\,\hat\varphi_{\alpha,m}(r)} \,\, ,
\label{eq:fermionicoperatoralphaandnolegindex}
\end{equation}
where $\hat\theta_{\alpha,m}(r)$ and $\hat\varphi_{\alpha,m}(r)$ are the phase and density bosonic fields, respectively, which obey the canonical commutation relations $[\hat\varphi_{\alpha,m}(r),\partial_{r'}\hat\theta_{\alpha',m'}(r')]=\delta_{\alpha,\alpha'}\,\delta_{m,m'}\,\delta(r-r')$, and $p$ is an odd integer. The pair operator, in the remapped lattice, reads either $\hat B^\dag_{r,m}=\hat c^\dag_{r,\uparrow,m}\hat c^\dag_{r,\downarrow,m}$ or $\hat B^\dag_{r,r+1,m}=\hat c^\dag_{r+1,\uparrow,m}\hat c^\dag_{r,\downarrow,m}$ (Fig.~\ref{fig:schemeoftheladderremapped}). We discuss now the bosonization form of $\hat B^\dag_{r,m}$.

By introducing \revtwo{the fields}
\begin{equation}
\hat\varphi_{\pm,m}=\frac{\hat\varphi_{\uparrow,m}\pm\hat\varphi_{\downarrow,m}}{\sqrt{2}} \qquad \hat\theta_{\pm,m}=\frac{\hat\theta_{\uparrow,m}\pm\hat\theta_{\downarrow,m}}{\sqrt{2}} \,\, ,
\end{equation}
the pair operator is bosonized using Eq.~\eqref{eq:fermionicoperatoralphaandnolegindex}:
\begin{eqnarray}
&&\hat B^\dag_{r,m}\sim e^{i\sqrt{2\pi}\,\hat\theta_{+,m}}\!\sum_{p}\left(e^{i2p\pi(n/2)r}\,e^{-ip\sqrt{2\pi}\,\hat\varphi_{+,m}}\right.\nonumber\\
&&\left.\hspace{3cm}+e^{-ip\sqrt{2\pi}\,\hat\varphi_{-,m}}\right) \,\, ,
\label{eq:bosonizationonsiteinteractionfermionicpairpp}
\end{eqnarray}
for odd $p$. Accordingly, the lowest non-oscillating harmonic of the NN interaction term now reads
\begin{eqnarray}
&&V\sum_r(\hat n_{r,\uparrow,\rev{m}}\hat n_{r,\downarrow,\rev{m}}+\hat n_{r,\downarrow,\rev{m}}\hat n_{r+1,\uparrow,\rev{m}})\nonumber\\
&&\hspace{2cm}\sim\int dr\,\cos\left(2\sqrt{2\pi}\,\hat\varphi_{-,\rev{m}}\right) \,\, .
\end{eqnarray}
When $V<0$, it pins the $\hat\varphi_{-,\rev{m}}$ field\rev{s}~\cite{giamarchi2003quantum} in Eq.~\eqref{eq:bosonizationonsiteinteractionfermionicpairpp} to $\hat\varphi_{-,\rev{m}}=0$, providing an effective $p=0$ harmonic. A further canonical transformation
\begin{equation}
\hat\theta_{+,m}=\frac{\hat\theta_{B,m}}{\sqrt{2}} \qquad \hat\varphi_{+,m}=\sqrt{2}\,\hat\varphi_{B,m} \,\, ,
\end{equation}
by introducing $q=2p$ and $n_B=n/2$, allows to recast Eq.~\eqref{eq:bosonizationonsiteinteractionfermionicpairpp} as
\begin{equation}
\hat B^\dag_{r,m}\sim e^{i\sqrt{\pi}\,\hat\theta_{B,m}(r)}\sum_{q}e^{iq\pi n_Br}\,e^{-iq\sqrt{\pi}\,\hat\varphi_{B,m}(r)} \,\, ,
\label{eq:bosonicoperatorb13}
\end{equation}
for $q$ even, therefore recovering a bosonic operator~\cite{giamarchi2003quantum,RevModPhys.83.1405}. An analogous result is found for $\hat B^\dag_{r,r+1,m}$. This result allows us to treat the pair as a single bosonic particle: $\hat B^\dag_{j,m}=\hat c^\dag_{j,m}\hat c^\dag_{j+1,m}\rightarrow \hat C^\dag_{z,m}$, localized at $z=j+1/2$, and therefore coarse-grain the system [Fig.~\ref{fig:schemeoftheladderremapped}(b)]. In the strongly-paired regime, a pair experiences a flux per plaquette equal to $2\Phi$.

The NNN interaction between fermions represent an intra-chain repulsive NN interaction $\tilde W$ between pairs. Moreover, even if the original fermions are not coupled by an inter-chain interaction, the presence of $t_\perp$, in addition to providing the inter-leg pair (Josephson) tunnelling
\begin{eqnarray}
&&\sum_ze^{-i2\Phi z}\hat C^\dag_{z,-\frac{1}{2}}\hat C_{z,+\frac{1}{2}}+{\rm H.c.}\nonumber\\
&&\hspace{0.5cm}\sim\int dz\,\cos\left[\sqrt{\pi}\left(\hat\theta_{B,+\frac{1}{2}}-\hat\theta_{B,-\frac{1}{2}}\right)+2\Phi z\right] \,\, ,
\end{eqnarray}
can perturbatively generate all interactions processes allowed by symmetry. The \rev{minimal} one that one expects is an on-site interaction between pairs on different legs
\begin{equation}
\sum_z\hat n_{z,-\frac{1}{2}}\hat n_{z,+\frac{1}{2}}\sim\int dz\,\cos\left[2\sqrt{\pi}\left(\hat\varphi_{B,+\frac{1}{2}}-\hat\varphi_{B,-\frac{1}{2}}\right)\right] \,\, .
\end{equation}
Therefore, we expect the system in the strongly-paired regime to be described \rev{by $\hat H_{\rm eff}\simeq\hat H^{(s)}+\hat H^{(a)}$, in terms of independent symmetric and anti-symmetric fields~\cite{PhysRevB.83.220518}
\begin{equation}
\hat\theta_{B,s/a}=\frac{\hat\theta_{B,+\frac{1}{2}}\pm\hat\theta_{B,-\frac{1}{2}}}{\sqrt{2}} \quad \hat\varphi_{B,s/a}=\frac{\hat\varphi_{B,+\frac{1}{2}}\pm\hat\varphi_{B,-\frac{1}{2}}}{\sqrt{2}} \,\, .
\end{equation}
\revtwo{Here,} $\hat H^{(s)}=\hat H^{(s)}_{\rm LL}$ is a gapless Luttinger liquid, whereas $\hat H^{(a)}$ is the self-dual sine-Gordon model~\cite{LECHEMINANT2002502}}
\begin{eqnarray}
&&\rev{\hat H^{(a)}=\hat H^{(a)}_{\rm LL}}+\tilde t_\perp\int dz\,\cos\left(\sqrt{2\pi}\,\hat\theta_{B,a}+2\Phi z\right)\nonumber\\
&&\hspace{2cm}+\tilde U\int dz\,\cos\left(2\sqrt{2\pi}\,\hat\varphi_{B,a}\right) \,\, .
\label{eq:pairedfermionshamiltonian}
\end{eqnarray}
\rev{This model belongs to the Ising universality class~\cite{LECHEMINANT2002502}, and} exhibits \rev{an Ising-type quantum phase transition}. We will now use this result in order to validate the emergence of bosons in the fermionic chain.

\begin{figure}[t!]
\centering
\includegraphics[width=8cm]{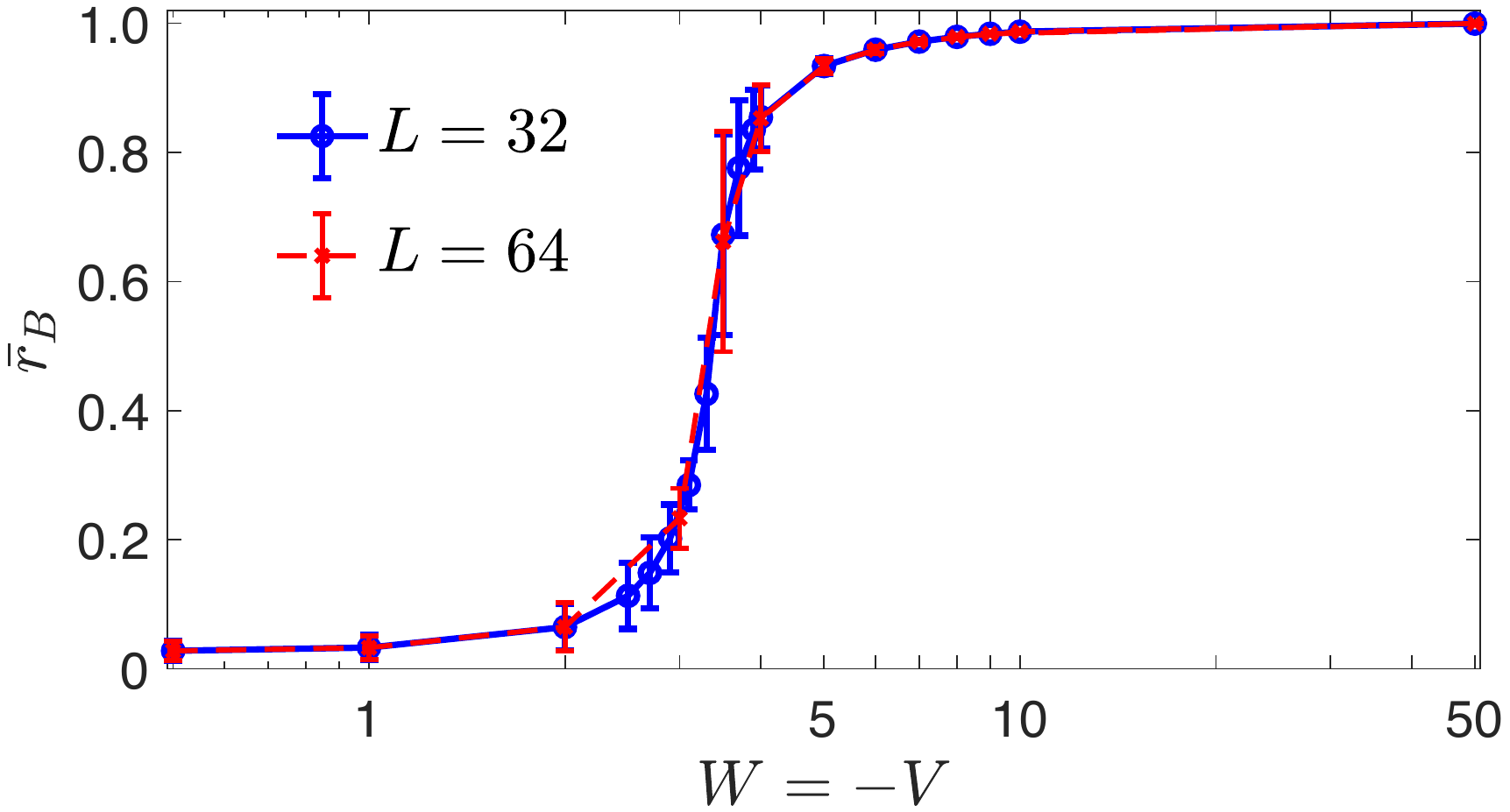}
\caption{Data of $\bar{r}_B$ (with uncertainty $\sigma_{\bar{r}_B}$) as a function of $W=-V$ (log-scale), for $L=32,64$ and $M=120$. The points are computed by simulating $\hat H_F$ for $k=41$ different values of $\Phi$, between $0$ and $2\pi$, and then $\bar{r}_B=k^{-1}\sum_\Phi{r_B}(\Phi)$ \revtwo{[see Eq.~\eqref{eq:parameterb}]}. The uncertainty is $\sigma_{\bar{r}_B}=(\max_\Phi r_B-\min_\Phi r_B)/2$.}
\label{fig:averagecooperdensity}
\end{figure}

\section{Numerical results}
\label{sec:numericalresults}
\revtwo{We now present our numerical results on the emergence of bosonic pairs in the system.} We simulate the Hamiltonian $\hat H_F$ by means of a DMRG algorithm that is the same used in Ref.~\cite{rossini2019anyonic}. For fixed values of $L$, $N$, $t_\perp$, $V$, $W$, and $\Phi$, after the initial infinite-DMRG sweep, a number of sweeps $S$ of finite-DMRG are performed in order to variationally find the density matrix of the system. During the sweeps, we truncate the dimension of the density matrix keeping up to $M$ states, \rev{where $M$} is chosen such that the truncation error does not exceed $\sim10^{-7}$~\cite{SCHOLLWOCK201196}. Because of the high numerical complexity of the problem, we can scan a limited range of parameters. Specifically, \revtwo{here, we present numerical data for} $t_\perp=0.3\,t$, $n=1/4$, and keep $W=-V>0$. We use $120\leq M\leq200$ and $3\leq S\leq 5$ depending on the observable that we measure, and on the value of $L$. \revtwo{We refer the interested reader to Appendix~\ref{sec:additionalnumericaldata} for additional numerical data.}

With the numerical algorithm that we use, we can measure only one- and two-point observables in terms of the original fermions $\hat c_{j,m}$, which means at most on-site observables for the emergent bosons $\hat B_{j,m}$. However, as we discuss below, the emergent bosonic physics can be detected already by looking at the pair density $n_{B,j,m}(\Phi)=\langle\Psi_{\rm GS}(\Phi)|\hat n_{B,j,m}|\Psi_{\rm GS}(\Phi)\rangle$, where $\hat n_{B,j,m}=\hat B^\dag_{j,m}\hat B_{j,m}$, which is the focus of the rest of our work \revtwo{(for a discussion on the measurement of the inter-leg current, the reader is referred to Appendix~\ref{sec:numericalresultscurrent})}.

\subsection{Detecting fermionic pairs}
A first evidence of the formation of pairs is provided by comparing the average local pair density $n_{B}(\Phi)=L^{-1}\sum_jn_{B,j,m}(\Phi)$ with the fermionic density $n$. Since we expect $n_B\ll n$ and $n_B=n/2$ [Eq.~\eqref{eq:bosonicoperatorb13}] in the unpaired and paired regimes, respectively, monitoring how \revtwo{the quantity}
\begin{equation}
r_B(\Phi)\coloneqq\frac{2n_B(\Phi)}{n} \,\, ,
\label{eq:parameterb}
\end{equation}
varies as $W$ is scanned from $W=0$ to large values provides information on the \rev{emergence of bosonic pairs in the system}.

\begin{figure}[t]
\centering
\includegraphics[width=8.5cm]{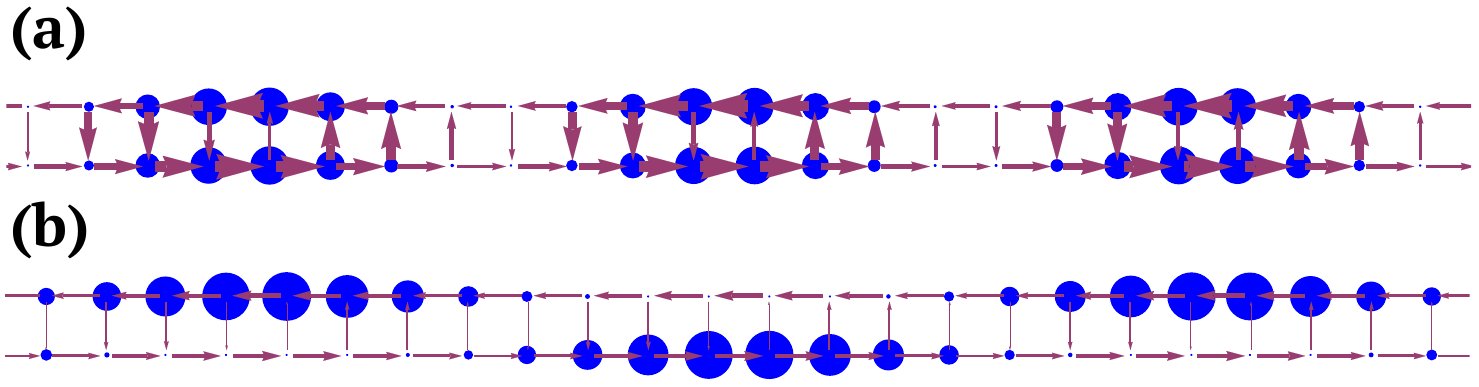}

\vspace{0.1cm}
\includegraphics[width=4.56cm]{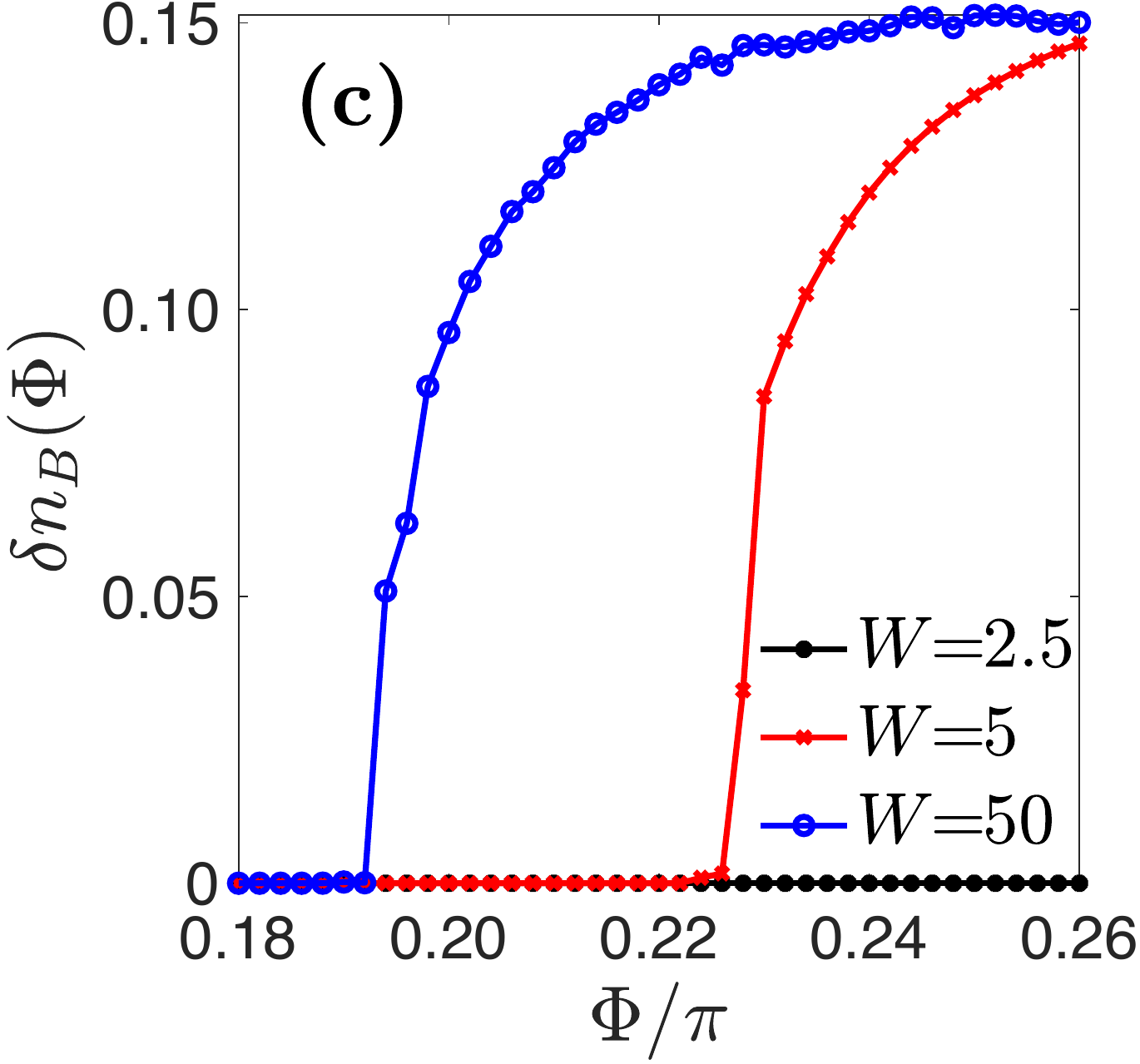}\!\!
\includegraphics[width=4.095cm]{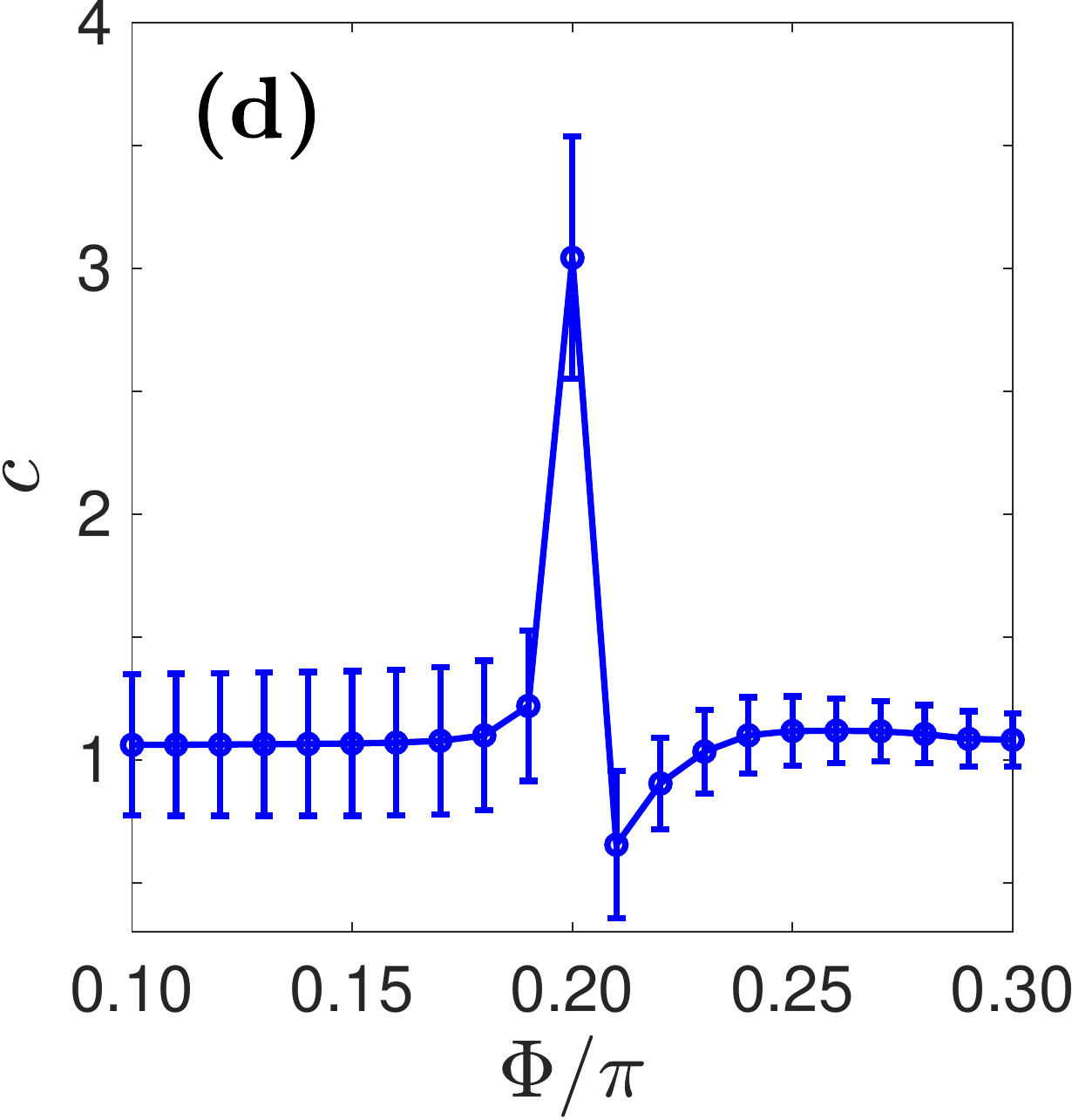}
\caption{Spatial configuration of the ladder [local density $\hat n_{B,j,m}$ (blue dots) and of fermionic intra- and inter-chain currents (arrows)] in the paired regime, identifying \textbf{(a)} a vortex density wave (VDW), and \textbf{(b)} a relative CDW. The flux drives a transition between these two phases as quantified in \textbf{(c)}: Data of $\delta n_B(\Phi)$ \revtwo{[see Eq.~\eqref{eq:expressionfordeltab}]} for $L=128$, $M=120$, and $W$ as in the legend. For $W=2.5$, no phase transition occurs. For $W=5,50$, the transition at a critical value $\Phi_c(W)$ is detected by a transition between $\delta n_B=0$ (VDW, $\Phi<\Phi_c$) and $\delta n_B\neq0$ (CDW, $\Phi>\Phi_c$). \textbf{(d)} Central charge $c$ for $W=50$, $L=96$, and $M=200$. Sufficiently far from the transition point $\Phi_c/\pi\simeq0.196$ [panel (c) and Fig.~\ref{fig:averagecooperdensity3}(b)], the fitted value of $c$ are consistent with $c=1$.}
\label{fig:averagecooperdensity2}
\end{figure}

The result is shown in Fig.~\ref{fig:averagecooperdensity}. We simulate $\hat H_F$ for $k=41$ values of $\Phi\in[0:2\pi]$, and show the flux average $\bar{r}_B=k^{-1}\sum_\Phi{r_B}(\Phi)$. We see that, for small $W$, $\bar{r}_B\simeq0$, whereas it approaches $\bar{r}_B=1$ as $W$ is increased. Between these two regimes, there is a wide range of $W$ in which $\bar{r}_B$ smoothly interpolates between $0$ and $1$. In such a region, fermions and bosonic pairs coexist, and the number of pairs $n_B(\Phi)$ is found to fluctuate with $\Phi$, quantified by the uncertainties on the data. For $W\gtrsim 5$, instead, such fluctuations are suppressed, and the system approaches the fully-paired regime. Because of the large number of flux values that we need for each $W$, we use $L=32,64$ in order to keep a reasonable computational complexity. Importantly, for the simulated values of $L$, the data are almost overlapped, \rev{and show} no finite-size scaling.

\subsection{Flux-driven Ising-type transition}
We now discuss the existence of an Ising-type transition driven by the gauge flux, in the paired regime. The first striking feature is that, in this regime, the system undergoes a flux-driven transition between a vortex density wave (VDW) ($\Phi<\Phi_c$) and relative charge density wave (CDW) ($\Phi>\Phi_c$), for some critical value $\Phi_c$ that depends on the system parameters. This manifests itself in the spatial patterns of $n_{B,j,m}$ and the local currents along the ladder, as in Fig.~\ref{fig:averagecooperdensity2}(a),(b), in which the fermionic intra- and inter-chain currents~\cite{PhysRevB.91.140406,strinati2017laughlin} (arrows) are shown together with $n_{B,j,m}$ (blue dots). For $\Phi<\Phi_c$, an ordered arrays of vortices appears (along with a vanishing relative density $\delta n_{B,j}=n_{B,j,-\frac{1}{2}}-n_{B,j,+\frac{1}{2}}$ for all $j$), which is compatible with the locking of the relative {\em phase} field $\hat\theta_{B,a}$~\cite{strinati2017laughlin}. Instead, for $\Phi>\Phi_c$, the relative density becomes periodically modulated, signalling a (staggered) CDW order (locking of the relative {\em charge} field $\hat\varphi_{B,a}$) \revtwo{(see also Appendix~\ref{sec:numericalresultscurrent})}.

This allows to focus on the local density imbalance between the two legs as a function of $\Phi$:
\begin{equation}
\delta n_{B,j}(\Phi)=\left|\langle\Psi_{\rm GS}(\Phi)|\left(\hat n_{B,j,-\frac{1}{2}}-\hat n_{B,j,+\frac{1}{2}}\right)|\Psi_{\rm GS}(\Phi)\rangle\right| \,\, ,
\label{eq:expressionfordeltab}
\end{equation}
in order to quantify the transition. Specifically, we compute the space average $\delta n_{B}(\Phi)=L^{-1}\sum_j\delta n_{B,j}(\Phi)$ scanning $\Phi$ through the transition. The key result is shown in Fig.~\ref{fig:averagecooperdensity2}(c). We compare the results in the paired regime ($W=5,50$) with those in the unpaired regime ($W=2.5$). As evident, no transition occurs for $W=2.5$ ($\delta n_{B}=0$ for all $\Phi$, signalling no density imbalance), whereas an increase of $\delta n_{B}$ around $\Phi_c$ is found for $W=5,50$ ($\delta n_{B}=0$ for $\Phi<\Phi_c$ and $\delta n_{B}>0$ for $\Phi>\Phi_c$, signalling the transition from the VDW to the CDW phase).

\begin{figure}[t]
\centering
\includegraphics[width=4.68cm]{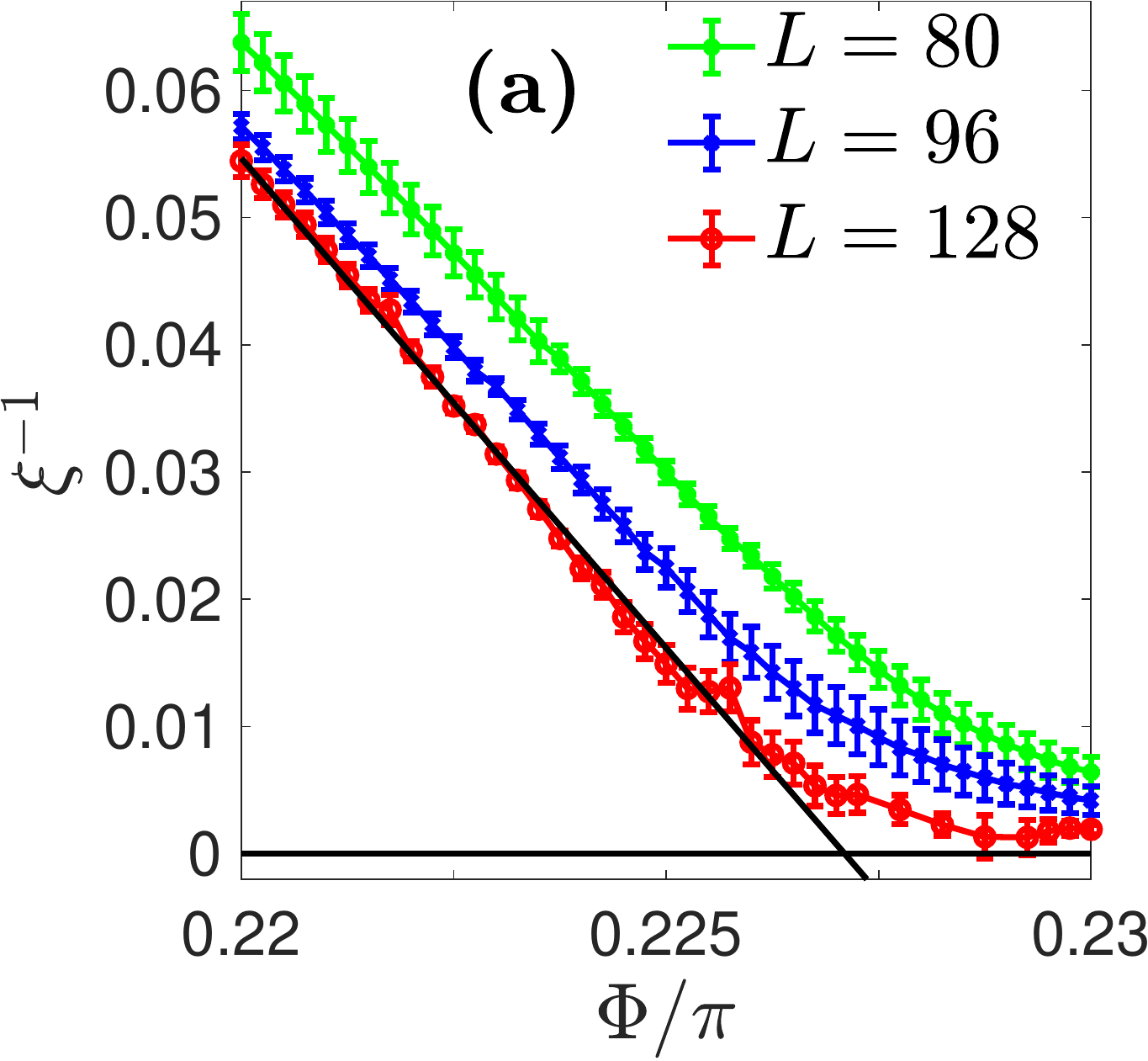}
\includegraphics[width=3.865cm]{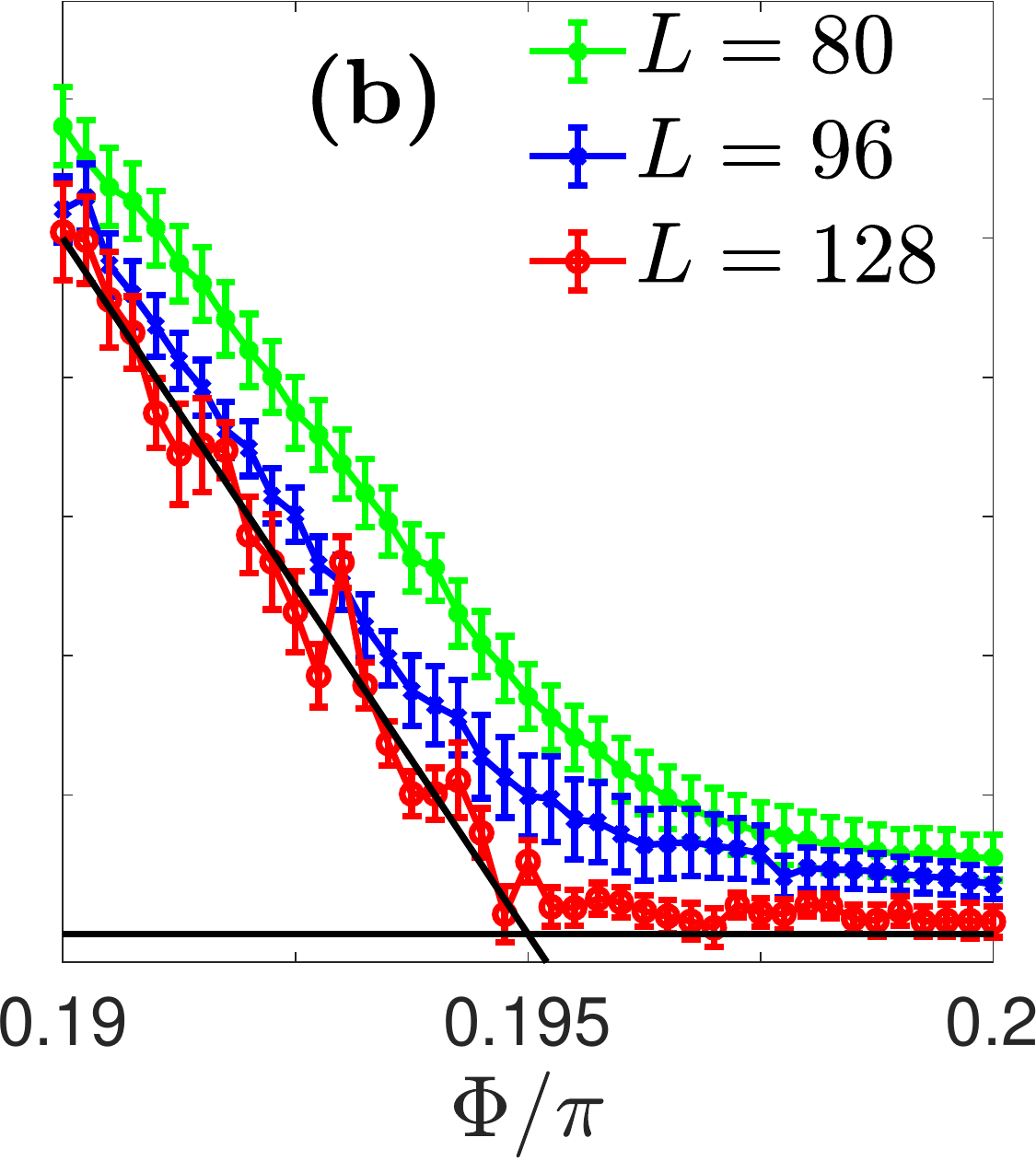}
\caption{Data for $\xi^{-1}(\Phi)$ from the simulation with localized impurity ($\mu=10^{-2}$), for $M=120$ and \textbf{(a)} $W=5$ and \textbf{(b)} $W=50$, and $L$ as in the legends. Black lines mark $\xi^{-1}=0$, and the linear fit to the data with $L=128$.}
\label{fig:averagecooperdensity3}
\end{figure}

We now try to illuminate the nature of this transition, and quantify the scaling of the correlation length $\xi$. This can be estimated by adding a localized impurity on one site $j_0$ of the $m=+1/2$ leg: $\hat H_{F,\mu}=\hat H_F-\mu\,\hat n_{j_0,+\frac{1}{2}}$, and analyzing the response of $\delta n_{B,j}$. The impurity locally enforces a density imbalance: if the GS is the VDW configuration, $\delta n_{B,j}$ is locally perturbed from the balanced configuration $\delta n_{B,j}=0$, but such a configuration is recovered after a characteristic length $\xi$: $\delta n_{B,j}\sim\delta n_{B,j_0}e^{-|j-j_0|/\xi}$. Instead, if the GS is a CDW configuration, the local imbalance forced by the impurity is preserved through the whole system ($\xi\rightarrow\infty$). By simulating $\hat H_{F,\mu}$, one can fit the data for  $\delta n_{B,j}$ \rev{vs. $j$} as $\Phi$ is varied across the transition and thus extract $\xi^{-1}(\Phi)\sim\Delta(\Phi)$, which is expected to exhibit scaling behaviour compatible \rev{with the Ising model in (1+1)-D}~\cite{PhysRevB.83.220518,gogolin2004bosonization,sachdev2001quantum}: $\Delta\sim|\Phi-\Phi_c|$ for $\Phi<\Phi_c$ and $\Delta=0$ for $\Phi>\Phi_c$.

The result of the simulation is shown in Fig.~\ref{fig:averagecooperdensity3}. We show $\xi^{-1}(\Phi)$ for $W=5,50$ and $L=80,96,128$. We observe that $\delta n_{B,j}$ exhibits an exponential decay, which is on top of spatial fluctuations (see also Fig.~\ref{fig:averagecooperdensity2}). In order to extract $\xi$ and account for such fluctuations, as well as finite-size effects, we fit the envelope of $\delta n_{B,j}$ with the function $f(j)=f_0e^{-|j-j_0|/\xi}$ (using $f_0$ and $\xi$ as fit parameters) three times, for $j\in[0.2\,L:L-\Delta L]$ and $\Delta L=0.15,0.2,0.3$. The resulting values of $\xi^{-1}$ and uncertainties are given by the average value and $(\max_{\Delta L}\xi^{-1}-\min_{\Delta L}\xi^{-1})/2$, respectively.
\rev{Within our numerical precision and limitations due to finite-size effects, our results are consistent with the linear closing of the gap $\Delta\sim|\Phi-\Phi_c|$, confirming the Ising transition.}

\subsection{Measuring the central charge}
A further observable to test the low-energy physics as in Eq.~\eqref{eq:pairedfermionshamiltonian} is given by the central charge $c$~\cite{giamarchi2003quantum}. A way to extract $c$ is to measure the von Neumann entropy (VNE), defined as $S_{\rm VNE}(\ell)=-{\rm Tr}[\hat\rho_\ell\ln(\hat\rho_\ell)]$, $\hat\rho_\ell$ being the reduced density matrix of a subpart of the system of size $\ell$, and fit it via the expression~\cite{Calabrese_2004}
\begin{equation}S(\ell)=a+\frac{c}{6}\,\ln\left[\left(\frac{2L}{\pi}\right)\sin\left(\frac{\pi\ell}{L}\right)\right] \,\, ,
\label{eq:calabreseformula}
\end{equation}
for OBC. Our numerical results for $W=50$ are shown in Fig.~\ref{fig:averagecooperdensity2}(d). In order to measure $c$ reliably, we use $M=200$, which significantly increases the computational time. We thus use $L=96$. Because of the fluctuating behaviour of the VNE, we extract the values of $c$ and relative uncertainties as in Refs.~\cite{strinati2017laughlin,Strinati_2018}. Away from the transition point, our data of $c$ are consistent with the value $c=1$ expected from Eq.~\eqref{eq:pairedfermionshamiltonian} (which contains a single gapless mode in the symmetric sector).

\section{Conclusions}
\label{sec:conclusions}
We analyzed the \rev{emergence of} bosonic pairs in the fermionic two-leg flux ladder \rev{with competing attractive and repulsive interaction}. We provided a phenomenological low-energy description in the strongly paired regime, which predicts the existence of an Ising-type transition between phases related by vortex-charge duality, and corroborated its validity by means of DMRG simulations. Although our numerics was limited to specific observables and values of parameters, due to the challenging numerical complexity of the problem, we observed a flux-driven Ising-type transition focusing on the divergence of the correlation length of the relative density order. Our work opens the possibility of creating interfaces in the flux-ladder between FQH and SC phases, thus opening a new intriguing path towards the possibility of hosting parafermions in flux-ladders~\cite{Clarke2013a,Lindner2012,vaezi2013fractional}. We leave these promising perspectives for future work.

\section*{Acknowledgements}
We thank Daniel Podolsky and Jonathan Ruhman for fruitful discussions. We are grateful to Davide Rossini for support. We acknowledge  support from the Israel Science Foundation (ISF), Grants No.~231/14 \rev{and 993/19} (E.~S. and M.~C.~S.) and No.~1452/14 (M.~C.~S.), and the U.S.-Israel Binational Science Foundation (BSF) Grant No.~2016130 and No.~2018726 (E.~S. and M.~C.~S.).

\vspace{0.2cm}
\appendix
\section{Direct evidence of the pairing gap}
\label{sec:directevidencepairinggap}

In Sec.~\ref{sec:lownenergytheoriinteractions}, we justify our bosonization treatment by the fact that, in the regime of parameters that we use, the presence of $V$ induces a hard pairing gap $\Delta_{\rm pair}$ between fermions, which corresponds to the energy that the system gains when two fermions bind to form a bosonic pair. In this appendix, we explicitly provide direct evidence of the presence of such a pairing gap.

\begin{figure}[t]
\centering
\includegraphics[width=4.4cm]{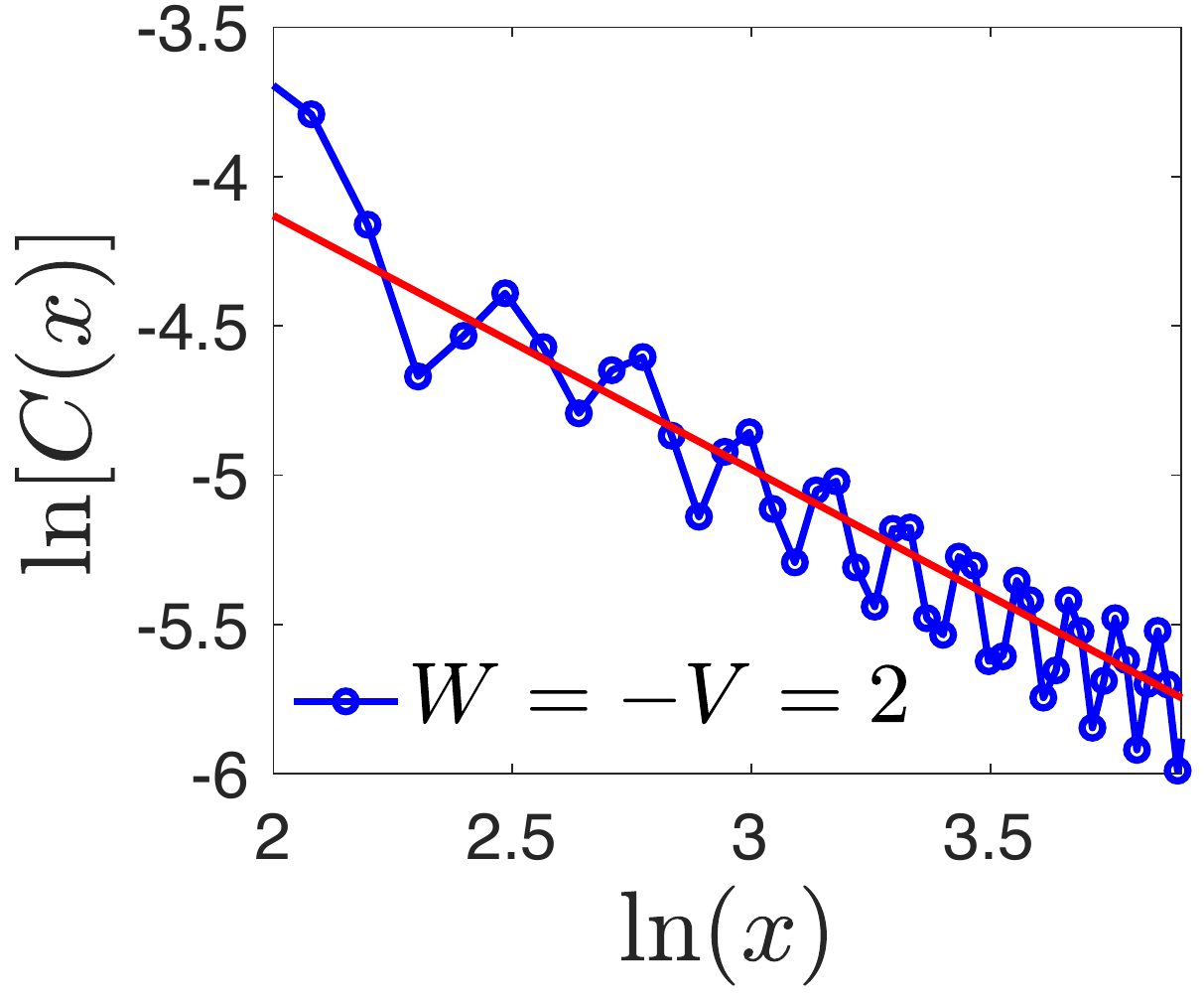}
\includegraphics[width=4.cm]{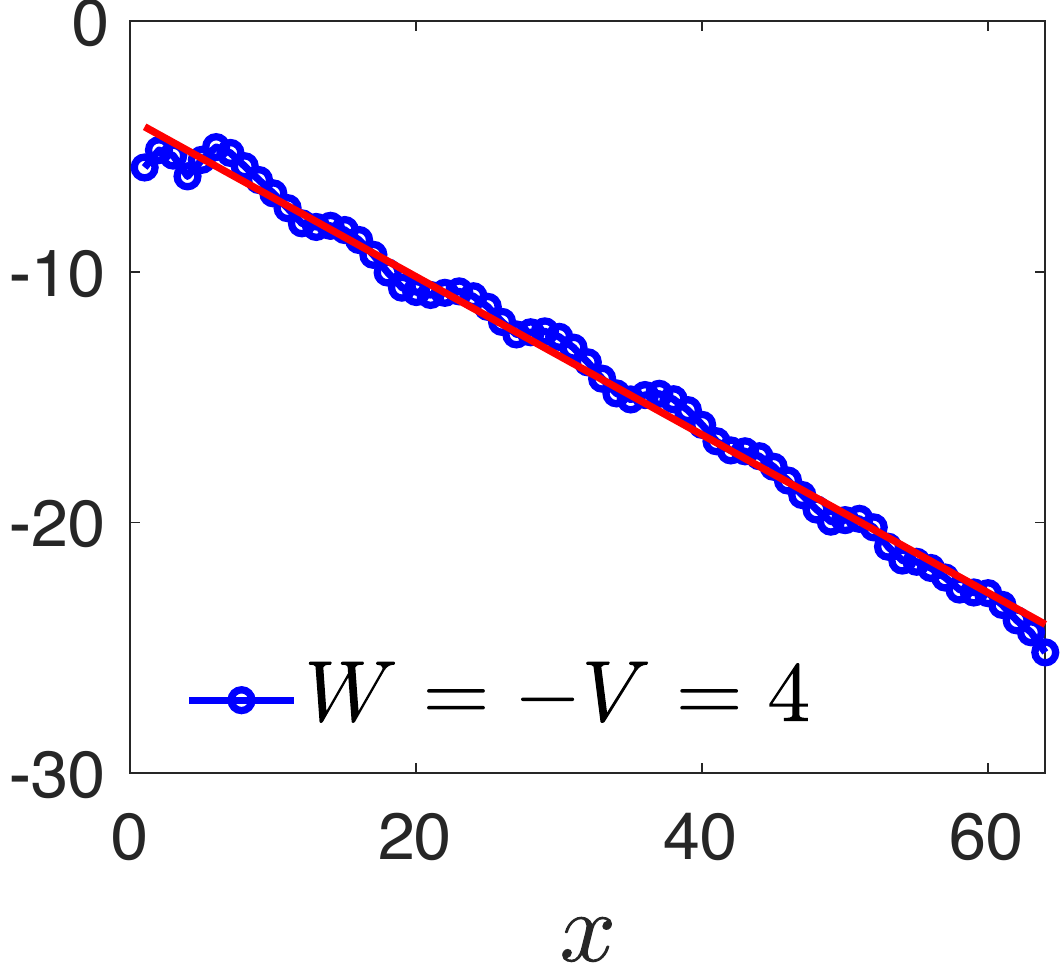}
\caption{Numerical results for $C(x)$ for the data in Fig.~\ref{fig:averagecooperdensity} ($L=64$ and $\Phi/\pi=0.3$). The data indeed show a power-law or exponential decay depending on the values of $V$. For this set of data, we find that for $|V|\lesssim3\,t$, $C(x)$ decays as a power law for sufficiently large $x$ (left panel, for $V=-2\,t$) highlighted by the linear fit in log-log scale (red line), while for $|V|\gtrsim4\,t$, $C(x)$ decays as an exponential for sufficiently large $x$ (right panel, for $V=-4\,t$) highlighted by the linear fit in log-linear scale (always given by the red line).}
\label{fig:correlationfunction}
\end{figure}

An observable that can monitor the presence of a pairing gap is the fermionic two-point correlation function
\begin{equation}
C(x)=\left|\langle\Psi_{\rm GS}|\hat c^\dag_x\hat c_0|\Psi_{\rm GS}\rangle\right| \,\, .
\label{eq:correlationfunctionpairinggap}
\end{equation}
The correlation function in Eq.~\eqref{eq:correlationfunctionpairinggap} was also used in different contexts, for example in Refs.~\cite{PhysRevB.96.085133,borla2019confinedphases}, in order to detect the occurrence of the pairing gap. At large $x$, $C(x)$ decays as a power law in the unpaired regime, and as an exponential in the paired regime, indeed due to the absence and presence of a pairing gap, respectively.

\begin{figure}[b]
\centering
\includegraphics[width=4.3cm]{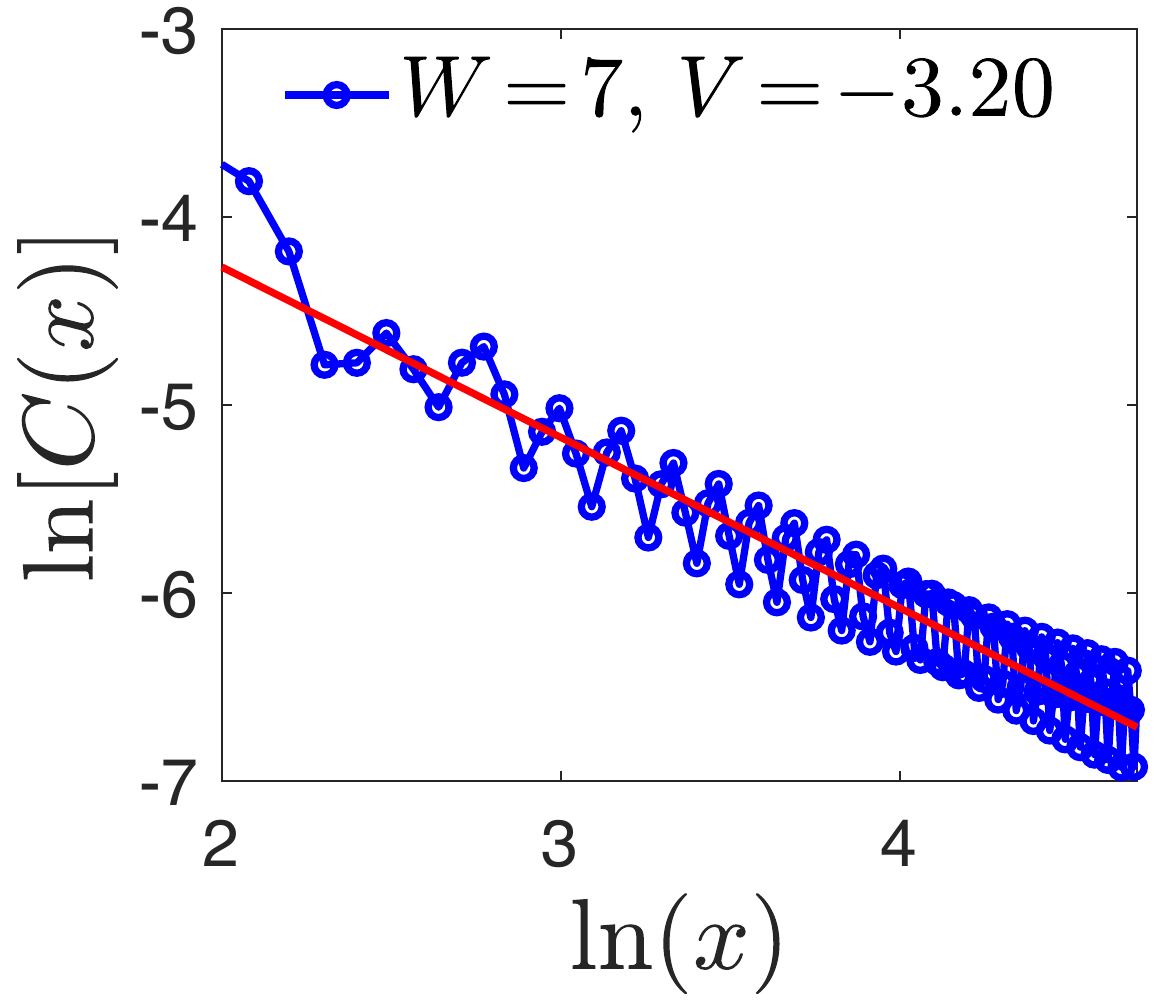}
\includegraphics[width=4.15cm]{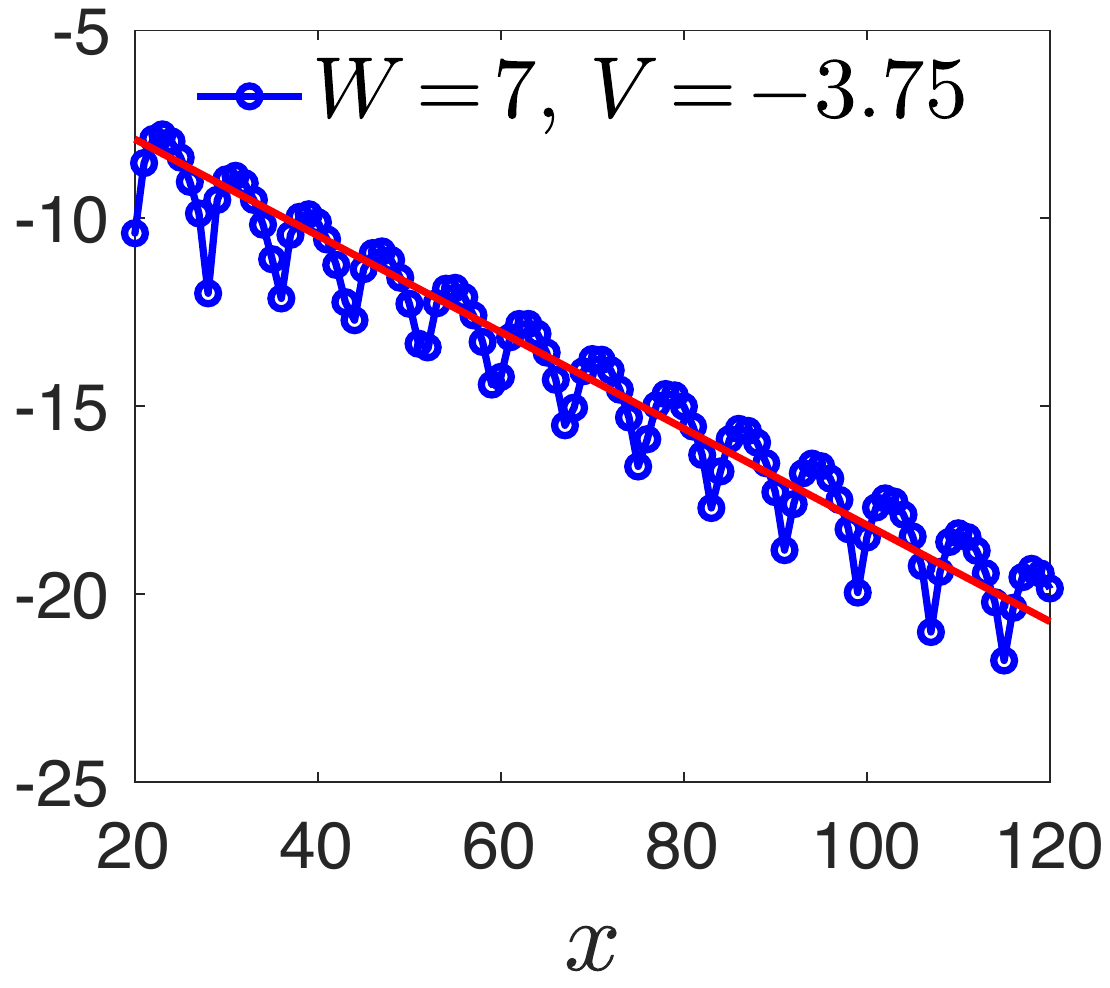}
\caption{Numerical results for $C(x)$ as in Fig.~\ref{fig:correlationfunction}. The data are those presented in Fig.~\ref{fig:datanormalizeddensity}, and are taken at $L=128$, $N=32$ (i.e., $n=1/4$), $t_\perp=0.3\,t$, $\Phi/\pi=0.3$, fixed $W=7\,t$ and (Left panel) $V=-3.20\,t$, and (Right panel) $V=-3.75\,t$.}
\label{fig:correlationfunction2}
\end{figure}

We find such a behaviour also from our numerical data. In particular, for the sake of clarity, we show in Fig.~\ref{fig:correlationfunction} the correlation function $C(x)$ for the numerical data for $L=64$ presented in Fig.~\ref{fig:averagecooperdensity}. By analyzing the long-distance behaviour of $C(x)$, we indeed see that, when $|V|$ is sufficiently small, $C(x)$ decays as a power-law, as can be appreciated by the linear decay in log-log scale (Fig.~\ref{fig:correlationfunction}, left panel), whereas when $|V|$ is sufficiently large, $C(x)$ decays as an exponential, $C(x)\sim e^{-\Delta_{\rm pair}x}$, as can be seen by the linear decay in log-linear scale (Fig.~\ref{fig:correlationfunction}, right panel). Specifically, we find that up to $|V|=3\,t$, $C(x)$ decays clearly as a power law, whereas for $|V|\geq4\,t$, it decays exponentially. The same results is found for the other simulations presented in this paper. Specifically, we show in Fig.~\ref{fig:correlationfunction2} the same analysis for the data in Fig.~\ref{fig:datanormalizeddensity} (Appendix~\ref{sec:additionalnumericaldata}), for which the simulation parameters are $L=128$, $N=32$ ($n=1/4$), $t_\perp=0.3\,t$, fixed $W=7\,t$.

This result tells us that, in the regime of parameters considered in Sec.~\ref{sec:numericalresults} (i.e., $|V|\geq 5\,t$), where the bosonic physics in which we are interested is discussed (the Ising VDW-CDW transition), the system is always in the strongly-paired regime, with the presence of a hard pairing gap. This also justifies our bosonization treatment in Sec.~\ref{sec:lownenergytheoriinteractions}.

\section{Additional numerical data}
\label{sec:additionalnumericaldata}
In this appendix, we show additional numerical data for different ranges of parameters, in order to show that the phenomenology that we discuss in our paper is not a consequence of a fine tuning of the system parameters.

\subsection{Varying the interaction strengths}
\begin{figure}[b]
\includegraphics[width=8cm]{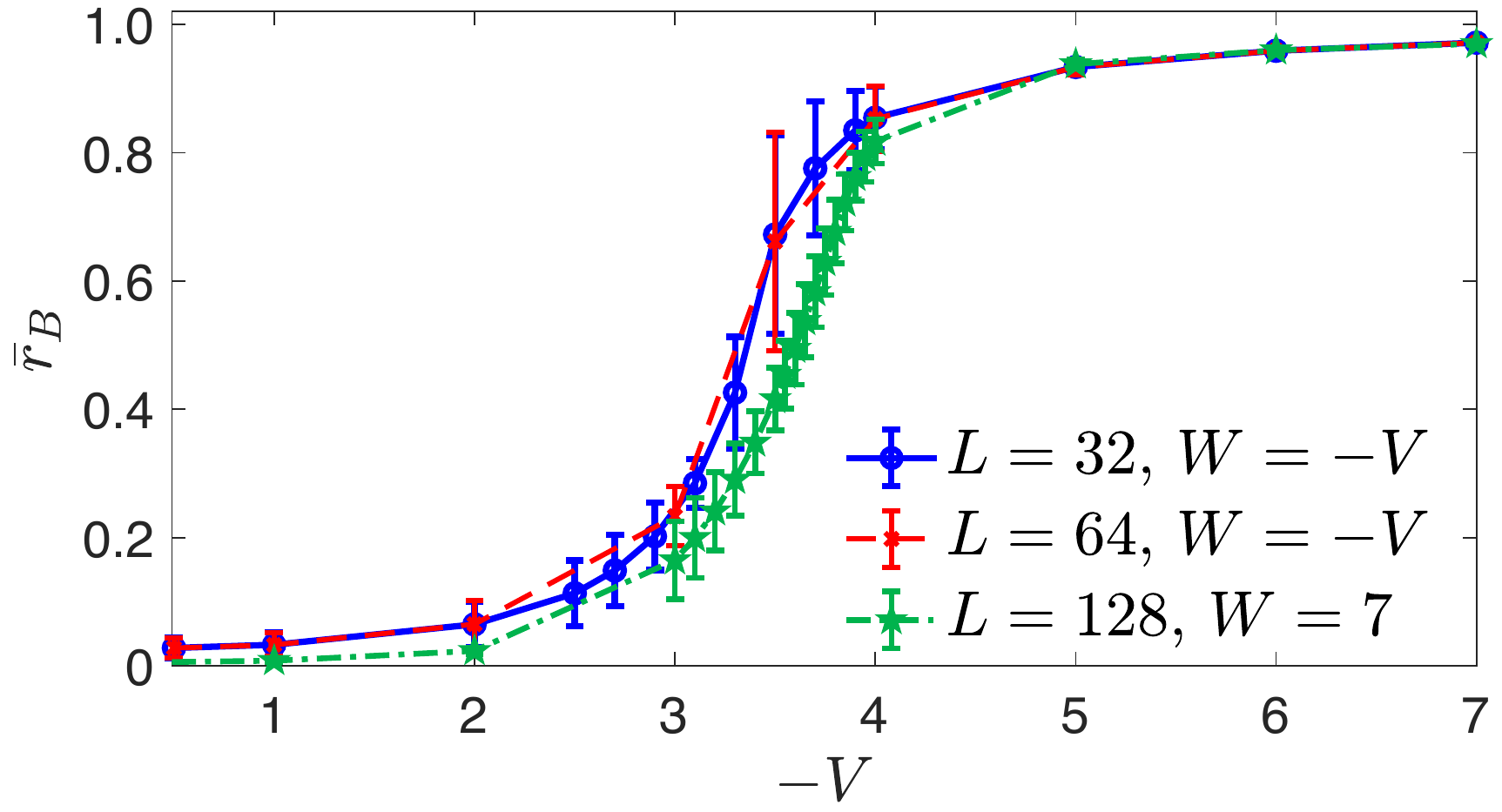}
\caption{Numerical data (green data) of $\bar r_B$ defined and computed as in Fig.~\ref{fig:averagecooperdensity}, for $L=128$, $N=32$ (i.e., $n=1/4$), $t_\perp=0.3\,t$, fixed $W=7\,t$ and by scanning $V$ from $V=0$ to $|V|=7\,t$. The data presented in Fig.~\ref{fig:averagecooperdensity} (blue and red data), which are computed for $L=32$ and $L=64$, and with $W=-V$, are superimposed to the new set of data for completeness.}
\label{fig:datanormalizeddensity}
\end{figure}

First, we relax the condition $W=-V$ considered in Sec.~\ref{sec:numericalresults}. We repeat the simulations as in Fig.~\ref{fig:averagecooperdensity}, but here we fix $W$, and vary $V$. In particular, we show in Fig.~\ref{fig:datanormalizeddensity} the result of a simulation keeping $W=7\,t$ and by scanning $V$ from $V=0$ to $|V|=7\, t$. We compute $\bar r_B$ as explained in Sec.~\ref{sec:numericalresults}. The other simulation parameters are: $L=128$, $N=32$ (i.e., $n=1/4$), $t_\perp=0.3\,t$, $M=120$ and $S=5$ sweeps. As we see, also in this other case, the quantity $\bar r_B$ displays a smooth increase from $\bar r_B\ll1$ (mostly unpaired fermions) for small $|V|$, to $\bar r_B\rightarrow 1$ (strongly-paired regime) as $|V|$ is increased.

\begin{figure}[t]
\centering
\includegraphics[width=8cm]{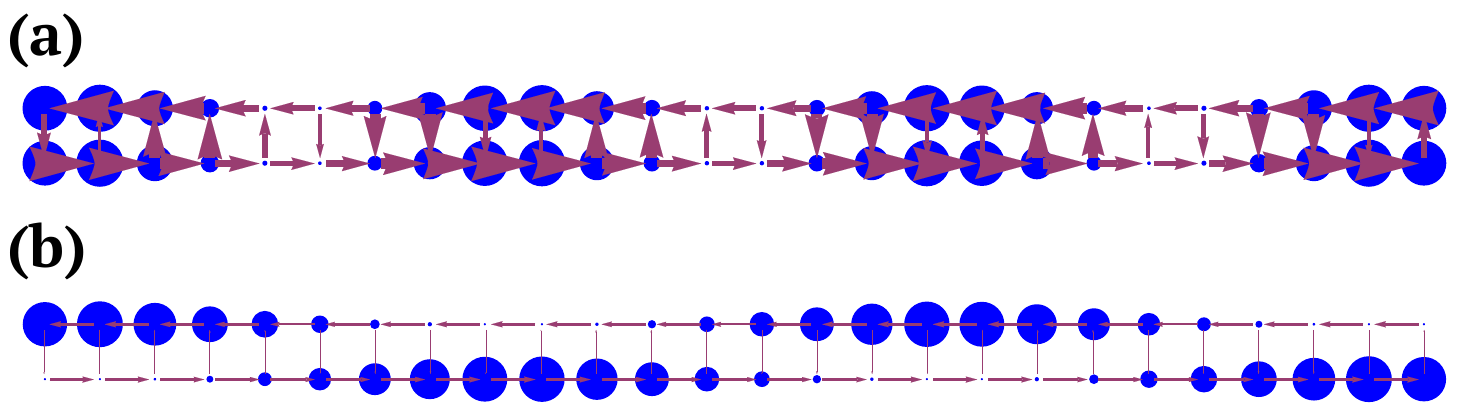}

\vspace{0.1cm}
\includegraphics[width=8cm]{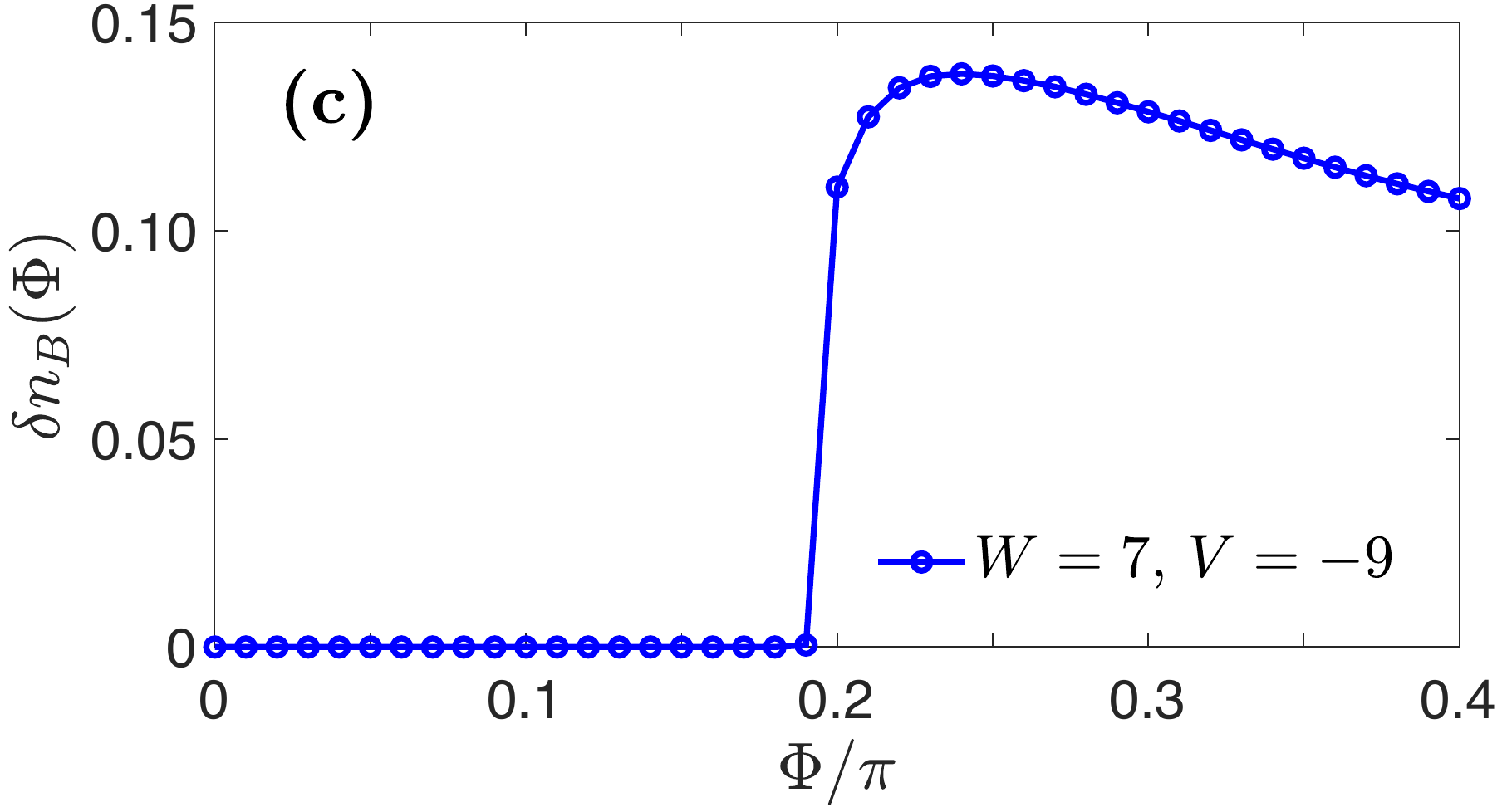}
\caption{Numerical results of the density and current configuration along the ladder, and $\delta n_B(\Phi)$ as in Fig.~\ref{fig:averagecooperdensity2}, panels \textbf{(a)}, \textbf{(b)} and \textbf{(c)}, here for the same data as in Fig.~\ref{fig:datanormalizeddensity} for $L=128$, specifically for $W=7\,t$ and $V=-9\,t$. The density and current pattern along the ladder is taken for \textbf{(a)} $\Phi/\pi=0.15$ (VDW), and \textbf{(b)} $\Phi/\pi=0.25$ (CDW).}
\label{fig:datanormalizeddensity2}
\end{figure}

\begin{figure}[t]
\includegraphics[width=8cm]{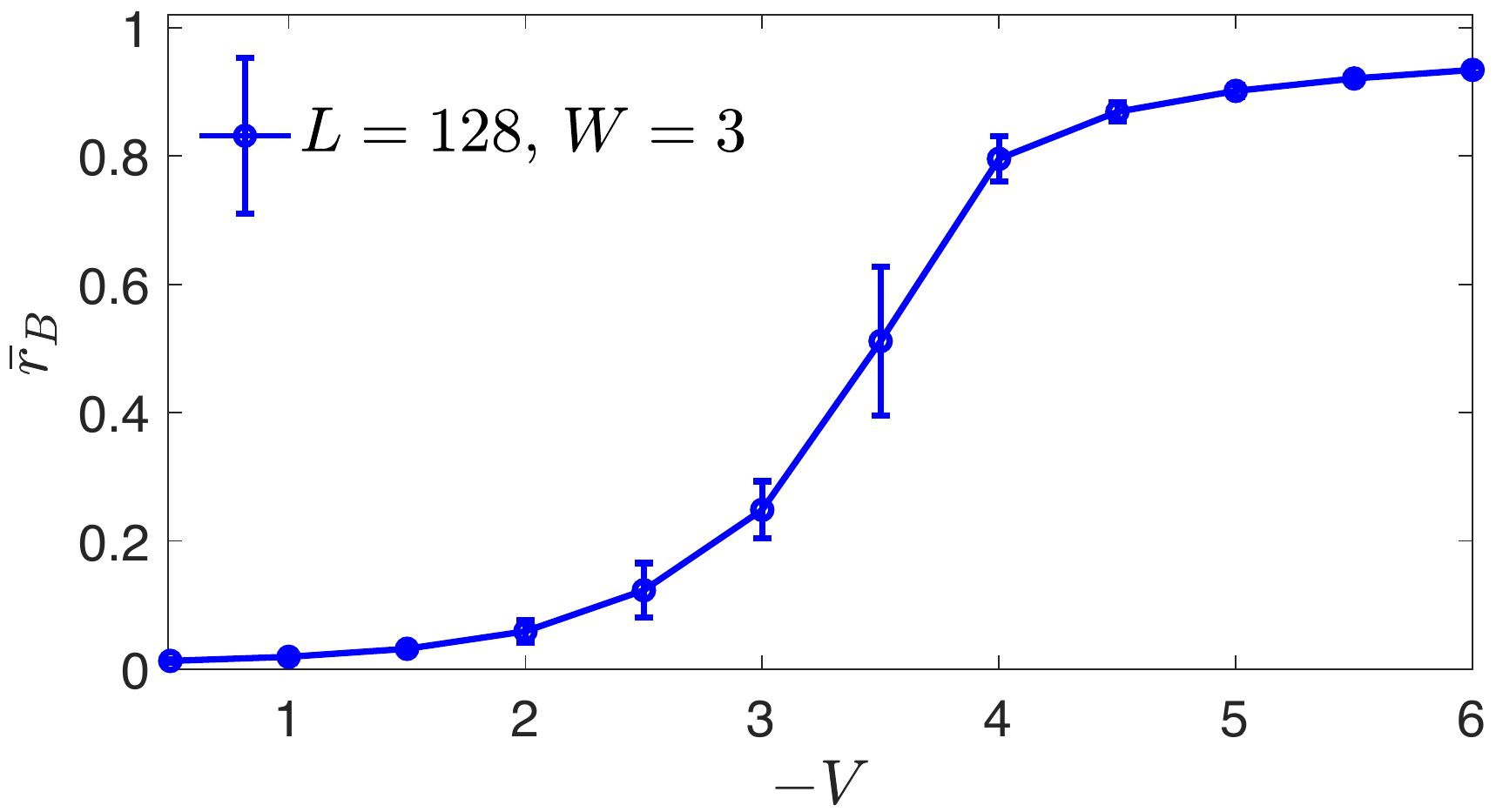}
\caption{Numerical data of $\bar r_B$ (see also Fig.~\ref{fig:datanormalizeddensity}), for $L=128$, $N=32$ (i.e., $n=1/4$), $t_\perp=0.3\,t$, fixed $W=3\,t$ and by scanning $V$ from $V=0$ to $|V|=6\,t$.}
\label{fig:datanormalizeddensity3}
\end{figure}

\begin{figure}[t]
\centering
\includegraphics[width=8cm]{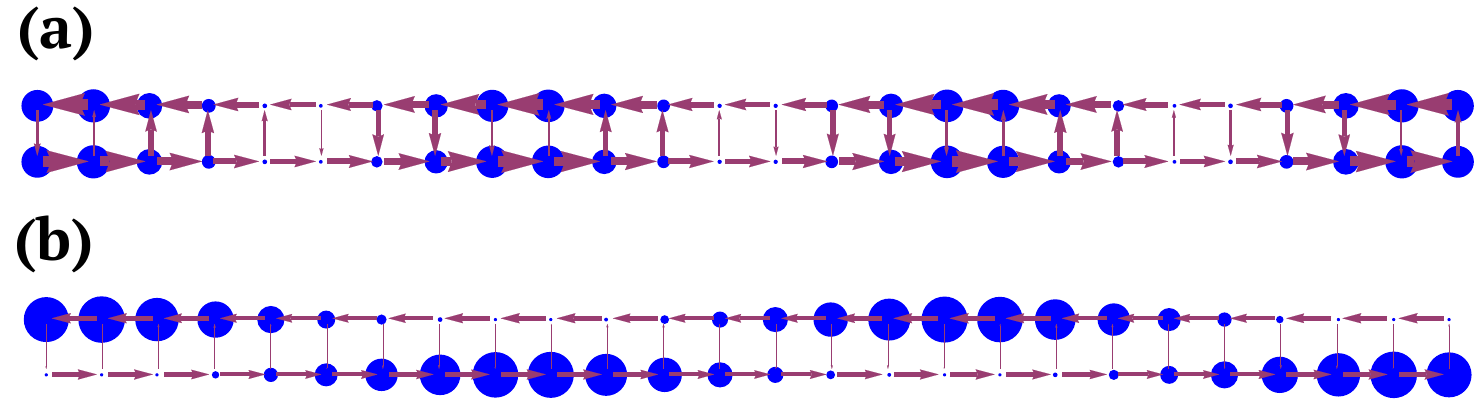}

\vspace{0.1cm}
\includegraphics[width=8cm]{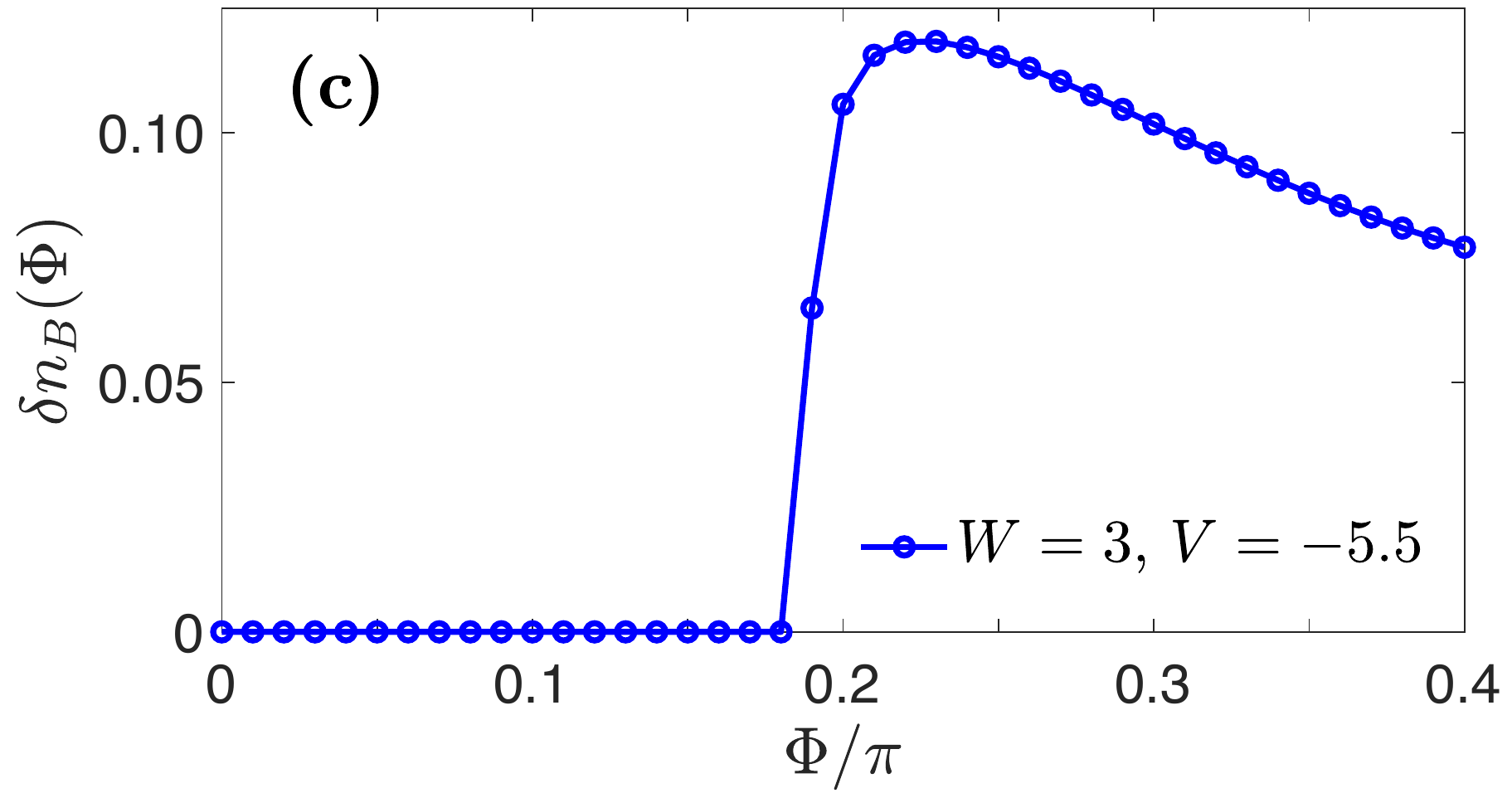}
\caption{Numerical results of the density and current configuration along the ladder, and $\delta n_B(\Phi)$, here for the same data as in Fig.~\ref{fig:datanormalizeddensity3}, specifically for $W=3\,t$ and $V=-5.5\,t$. The density and current pattern along the ladder is taken for \textbf{(a)} $\Phi/\pi=0.15$ (VDW), and \textbf{(b)} $\Phi/\pi=0.22$ (CDW).}
\label{fig:datanormalizeddensity4}
\end{figure}

We further show the data for $\delta n_B(\Phi)$, and the density and current configuration along the ladder in Fig.~\ref{fig:datanormalizeddensity2}, as in Fig.~\ref{fig:averagecooperdensity2}. Specifically, the data are shown for $W=7\,t$ and $V=-9\,t$. In panels \textbf{(a)} and \textbf{(b)}, we show the ladder configuration for $\Phi/\pi=0.15$ (in the VDW phase) and for $\Phi/\pi=0.25$ (in the CDW phase), respectively. We see that, apart from specific quantitative details, the same phenomenology discussed throughout the paper for $W=-V$ arises also for this other choice of $V$ and $W$.

The same numerical simulations as in Figs.~\ref{fig:datanormalizeddensity} and~\ref{fig:datanormalizeddensity2} are repeated for a different value of $W$, namely, we fix $W=3\,t$ and scan $|V|$ from $|V|=0$ to $|V|=6\,t$, since we observed the formation of clusters for larger values of $|V|$. The result of the simulation is shown in Figs.~\ref{fig:datanormalizeddensity3} and~\ref{fig:datanormalizeddensity4}. Again, the data of $\bar r_B$ in Fig.~\ref{fig:datanormalizeddensity3} show a smooth increase from almost zero to $\bar r_B\rightarrow1$ for sufficiently strong $|V|$. In Fig.~\ref{fig:datanormalizeddensity4}, $\delta n_B(\Phi)$ for $|V|=5.5\,t$ is shown, together with the density and current pattern configuration along the ladder, displaying once again the same phenomenology (VDW-CDW transition).

\begin{figure}[t]
\centering
\includegraphics[width=8cm]{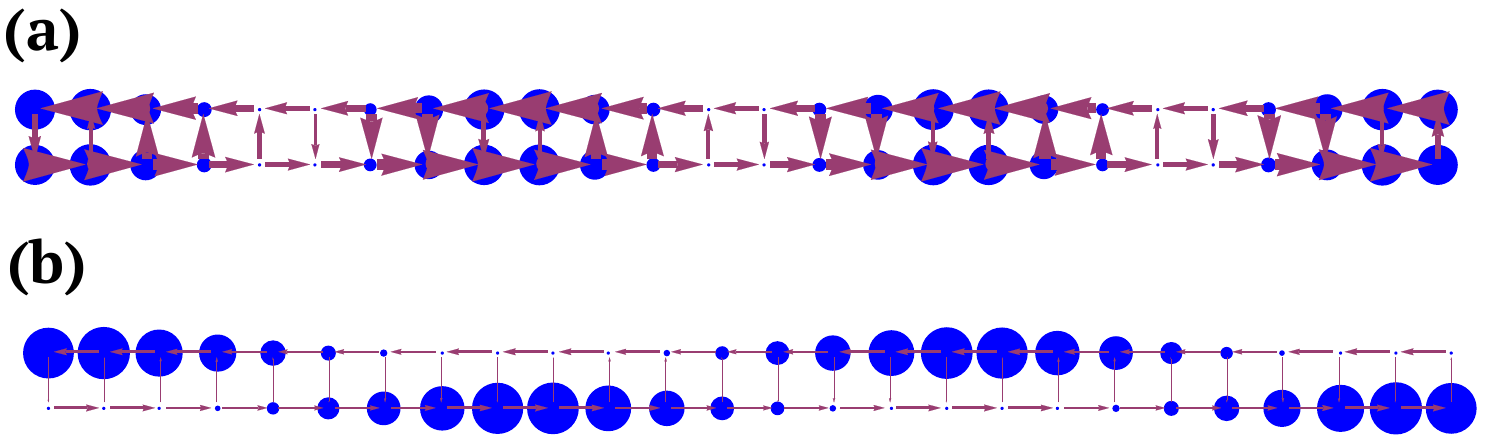}

\vspace{0.1cm}
\includegraphics[width=8cm]{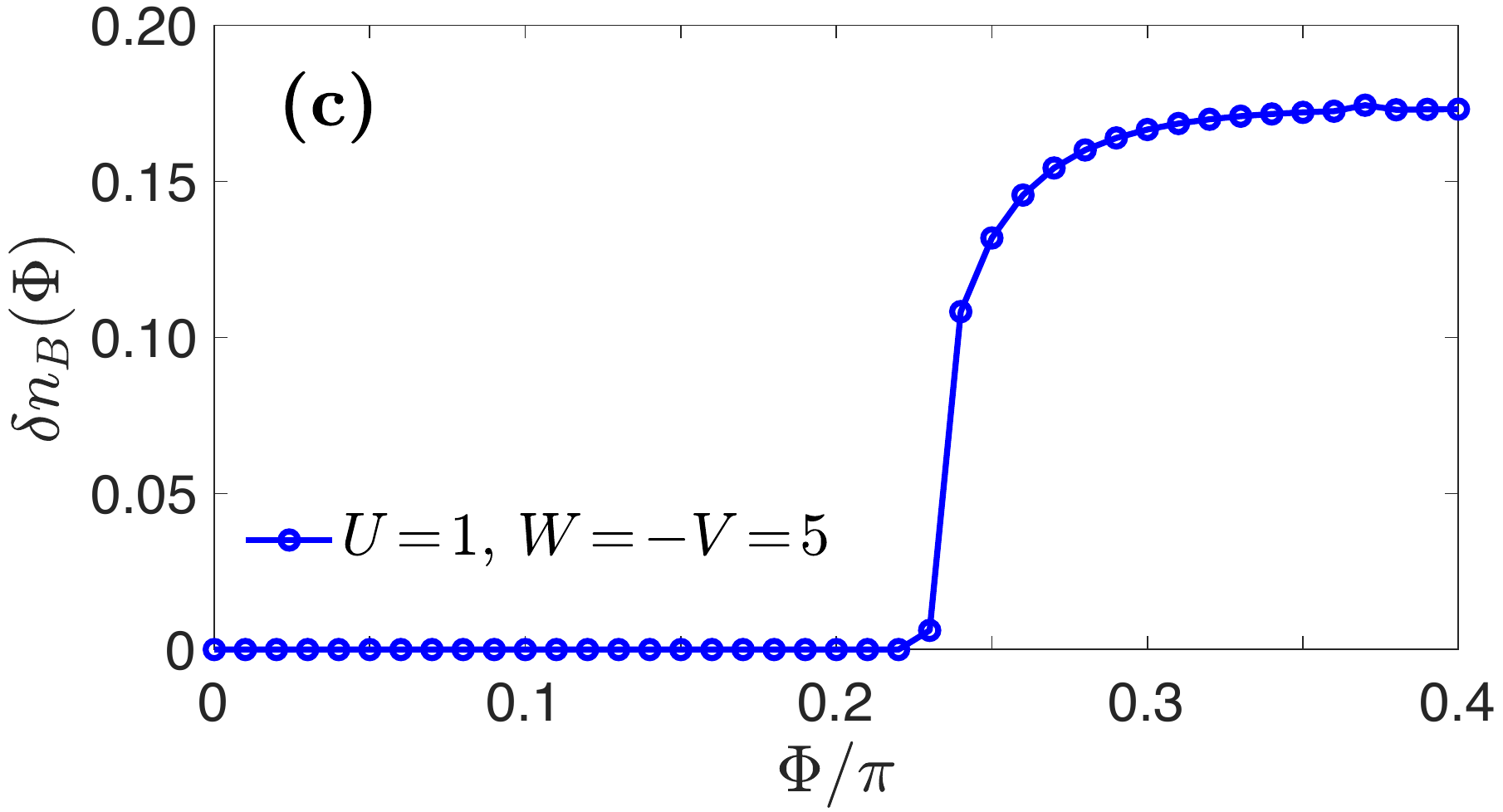}
\caption{Numerical results of the density and current configuration along the ladder, and $\delta n_B(\Phi)$, using the interaction Hamiltonian as in Eq.~\eqref{eq:interactionhamiltonian} with the inclusion of the term $\hat H_U$ [Eq.~\eqref{eq:interactionhamiltonianonsite}], with $U=1\,t$. The numerical parameters are $L=128$ and $N=32$, which is $n=1/4$, $W=-V=5\,t$, $t_\perp=0.3\,t$, $M=120$ and $S=5$ sweeps. The density and current configuration along the ladder for panels \textbf{(a)} and \textbf{(b)} is taken at $\Phi/\pi=0.15$ and $\Phi/\pi=0.35$, respectively.}
\label{fig:datanormalizeddensity5}
\end{figure}

\begin{figure}[t]
\centering
\includegraphics[width=8cm]{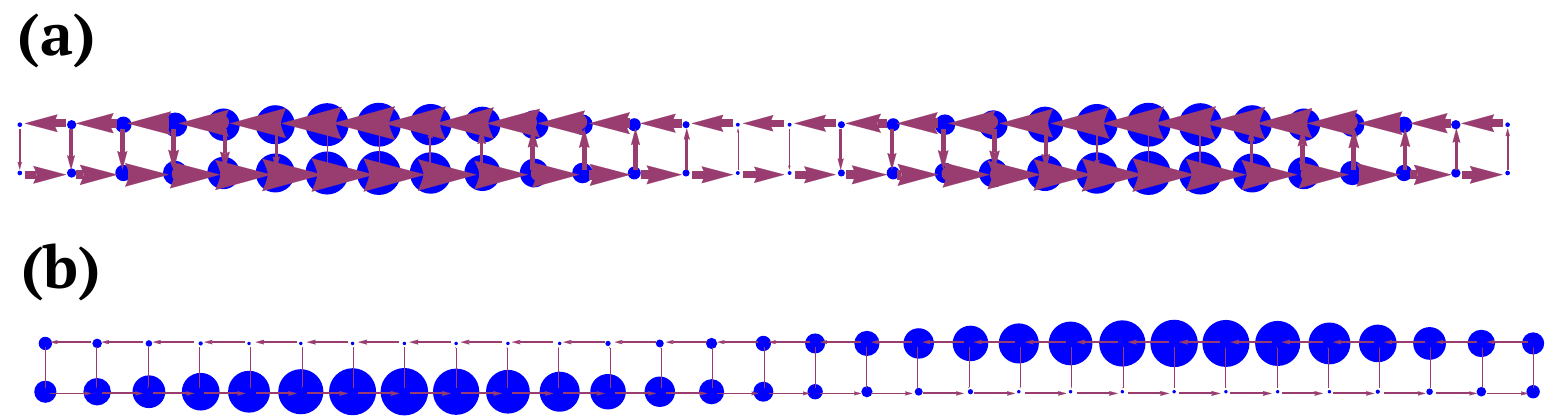}

\vspace{0.1cm}
\includegraphics[width=8cm]{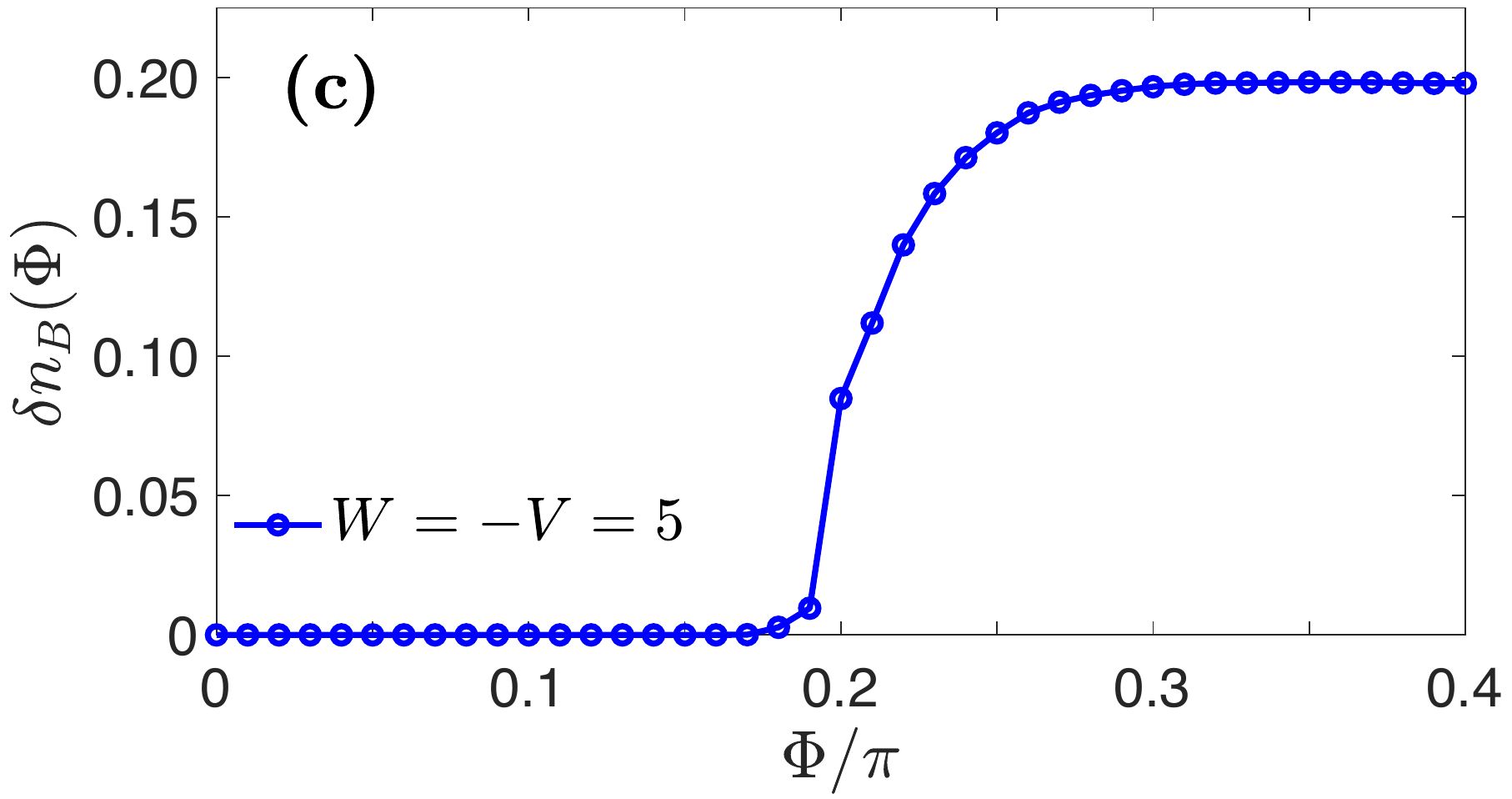}
\caption{Numerical results of the density and current configuration along the ladder, and $\delta n_B(\Phi)$, here for $L=160$ and $N=20$, which is $n=1/8$. The other numerical parameters are $W=-V=5\,t$, $t_\perp=0.3\,t$, $M=120$ and $S=3$ sweeps. The density and current pattern along the ladder is taken for \textbf{(a)} $\Phi/\pi=0.15$, and \textbf{(b)} $\Phi/\pi=0.25$.}
\label{fig:datanormalizeddensity6}
\end{figure}

\begin{figure}[t]
\includegraphics[width=8cm]{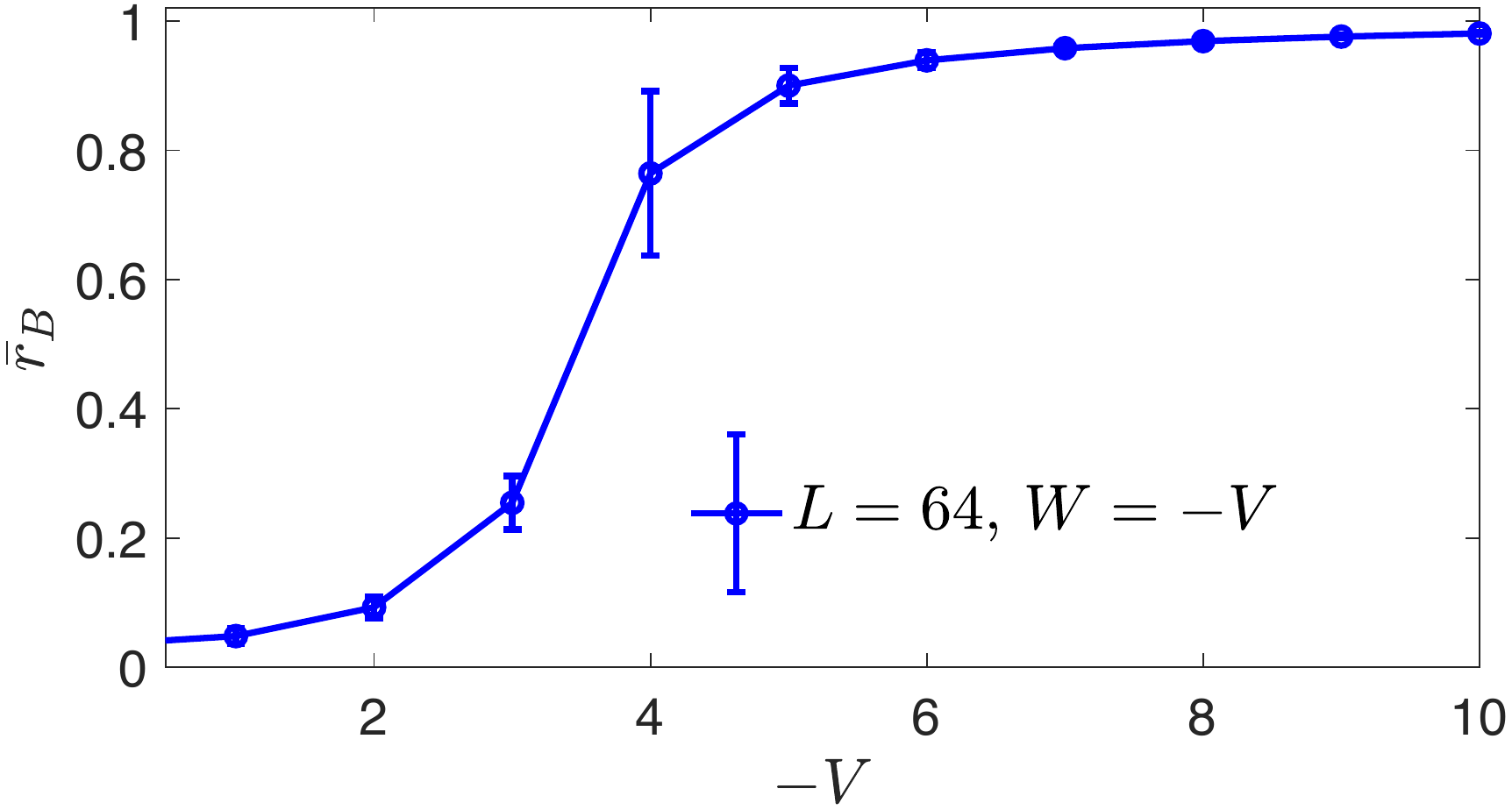}
\caption{Numerical data of $\bar r_B$ for $L=64$, $N=12$ (i.e., $n=1/4$), $t_\perp=0.5\,t$, and by scanning $V=-W$ from $V=0$ to $|V|=10\,t$.}
\label{fig:datanormalizeddensity7}
\end{figure}

\begin{figure}[t]
\centering
\includegraphics[width=8cm]{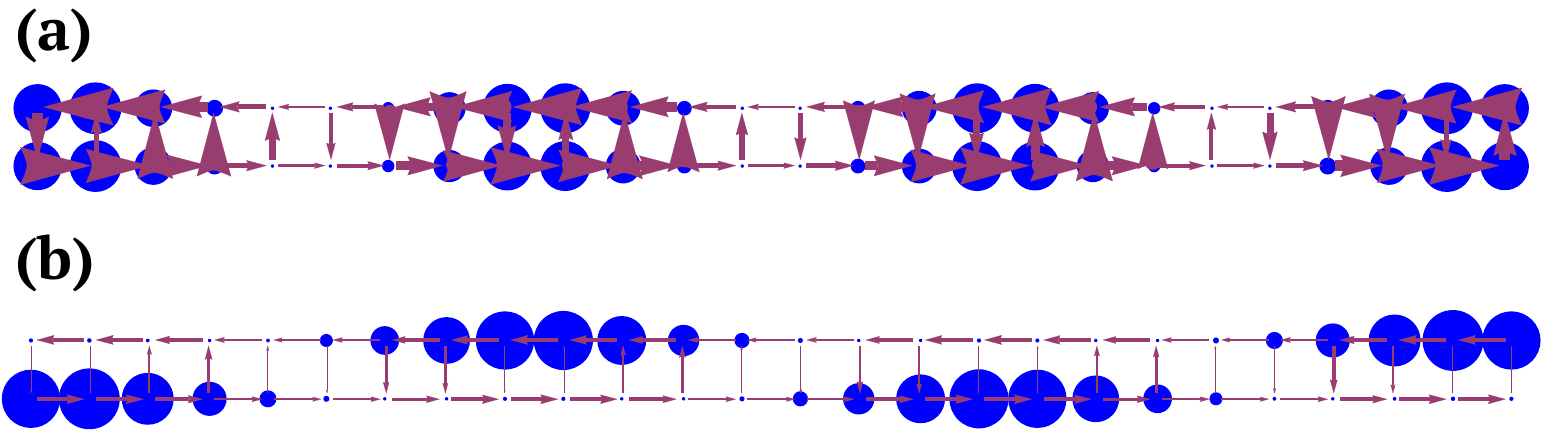}

\vspace{0.1cm}
\includegraphics[width=8cm]{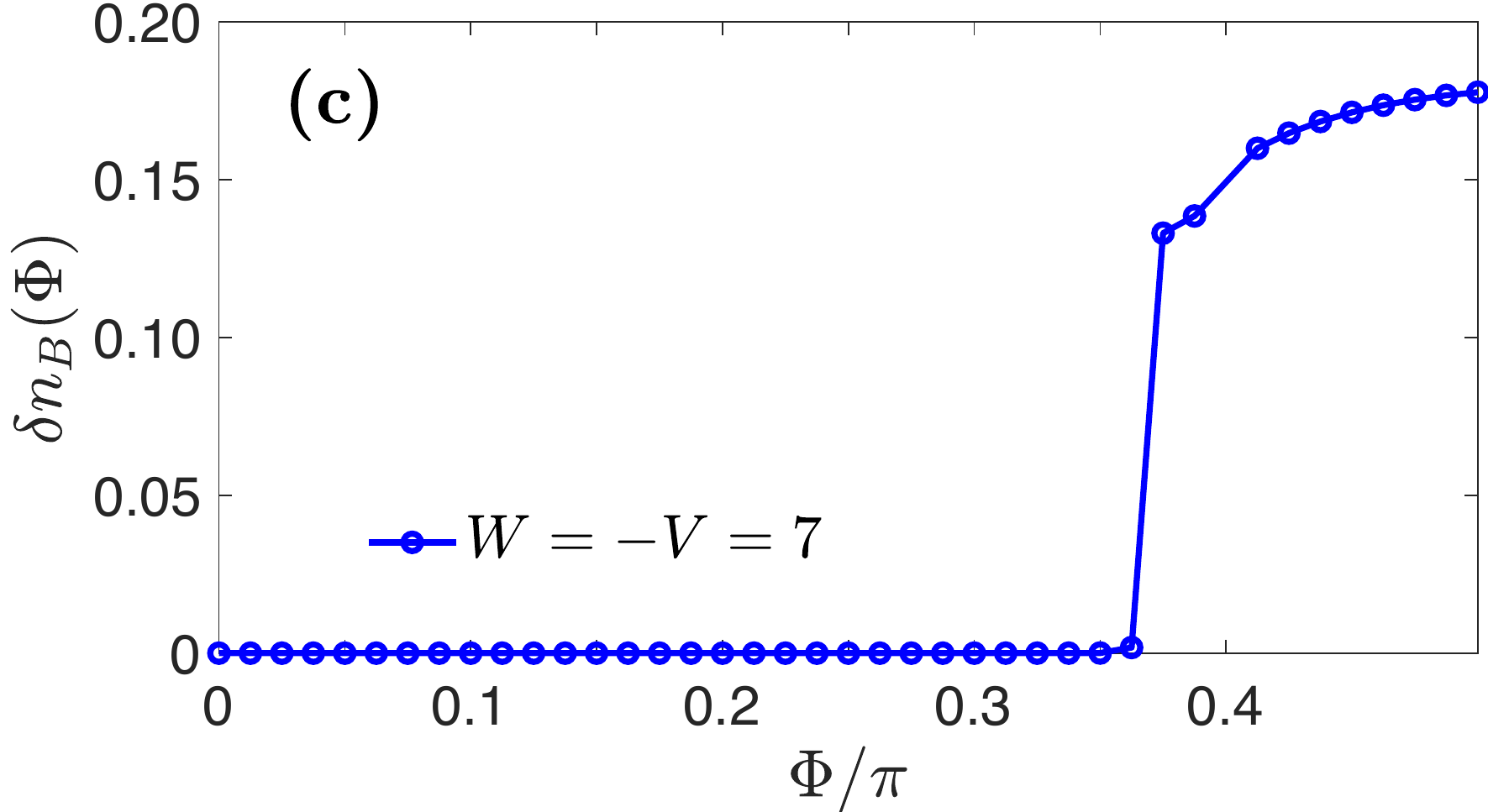}
\caption{Numerical results of the density and current configuration along the ladder, and density difference $\delta n_B(\Phi)$, here for the same data as in Fig.~\ref{fig:datanormalizeddensity7}, specifically for $W=-V=7\,t$. The density and current pattern along the ladder is taken for \textbf{(a)} $\Phi/\pi=0.25$ (VDW), and \textbf{(b)} $\Phi/\pi=0.45$ (CDW).}
\label{fig:datanormalizeddensity8}
\end{figure}

We now discuss the presence of an inter-leg density-density interaction of the form
\begin{equation}
\hat H_U=U\sum_j\hat n_{j,-\frac{1}{2}}\hat n_{j,+,\frac{1}{2}} \,\, ,
\label{eq:interactionhamiltonianonsite}
\end{equation}
at the level of the original microscopic Hamiltonian. The original microscopic model (Sec.~\ref{sec:model}) does not include such a term, but as we say in Sec.~\ref{sec:lownenergytheoriinteractions}, such an interaction term is generated in the effective model by $t_\perp$, $V$ and $W$. We show in Fig.~\ref{fig:datanormalizeddensity5} the result of a simulation using the interaction Hamiltonian $\hat H_{\rm int}$ as in Eq.~\eqref{eq:interactionhamiltonian}, with the inclusion of the on-site interaction term as in Eq.~\eqref{eq:interactionhamiltonianonsite}, with $U=1\,t$. As we see, the presence of $U>0$ does not dramatically affect the physics that we discussed throughout the paper.

\begin{figure}[t]
\centering
\includegraphics[width=4.45cm]{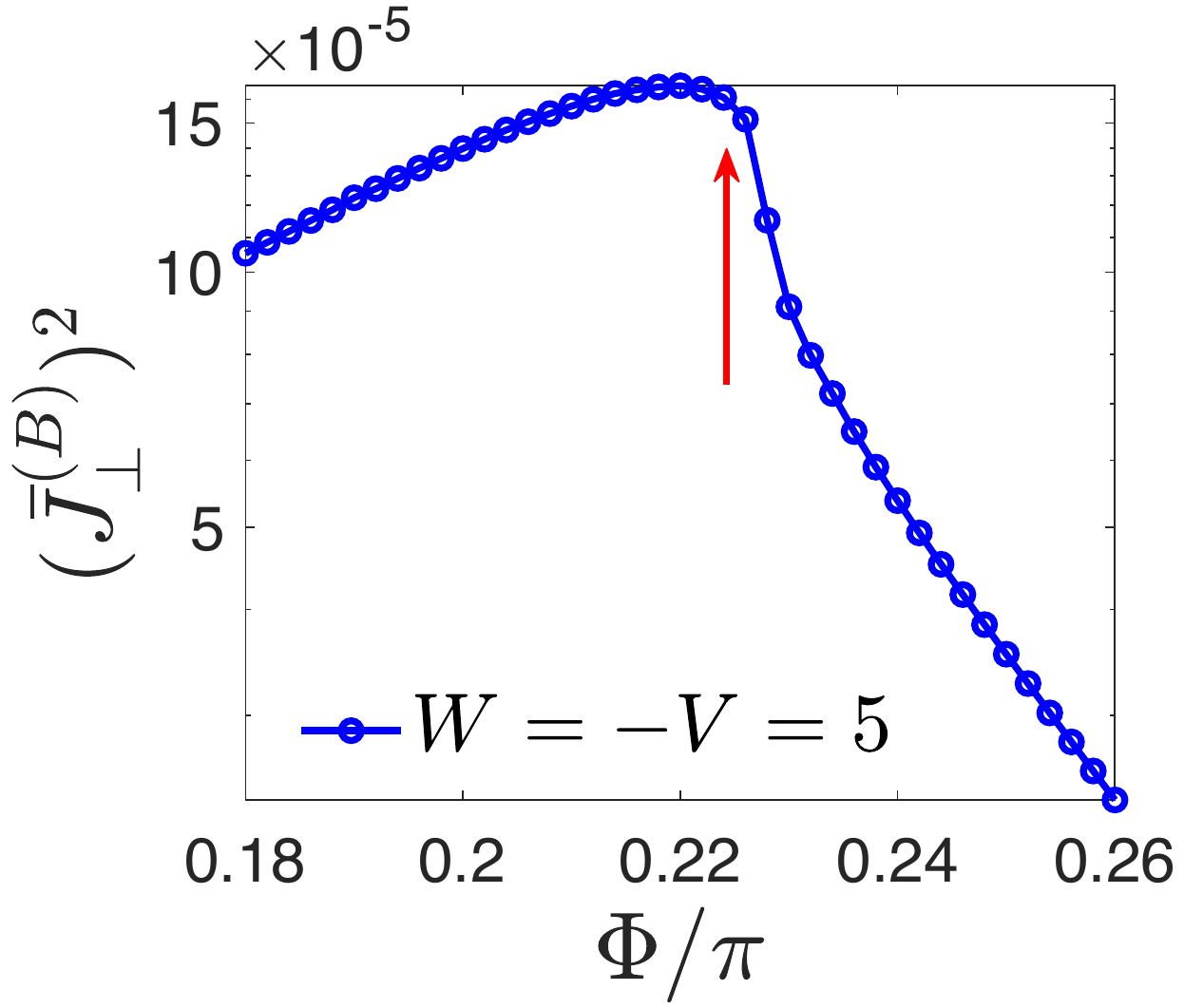}
\includegraphics[width=4.1cm]{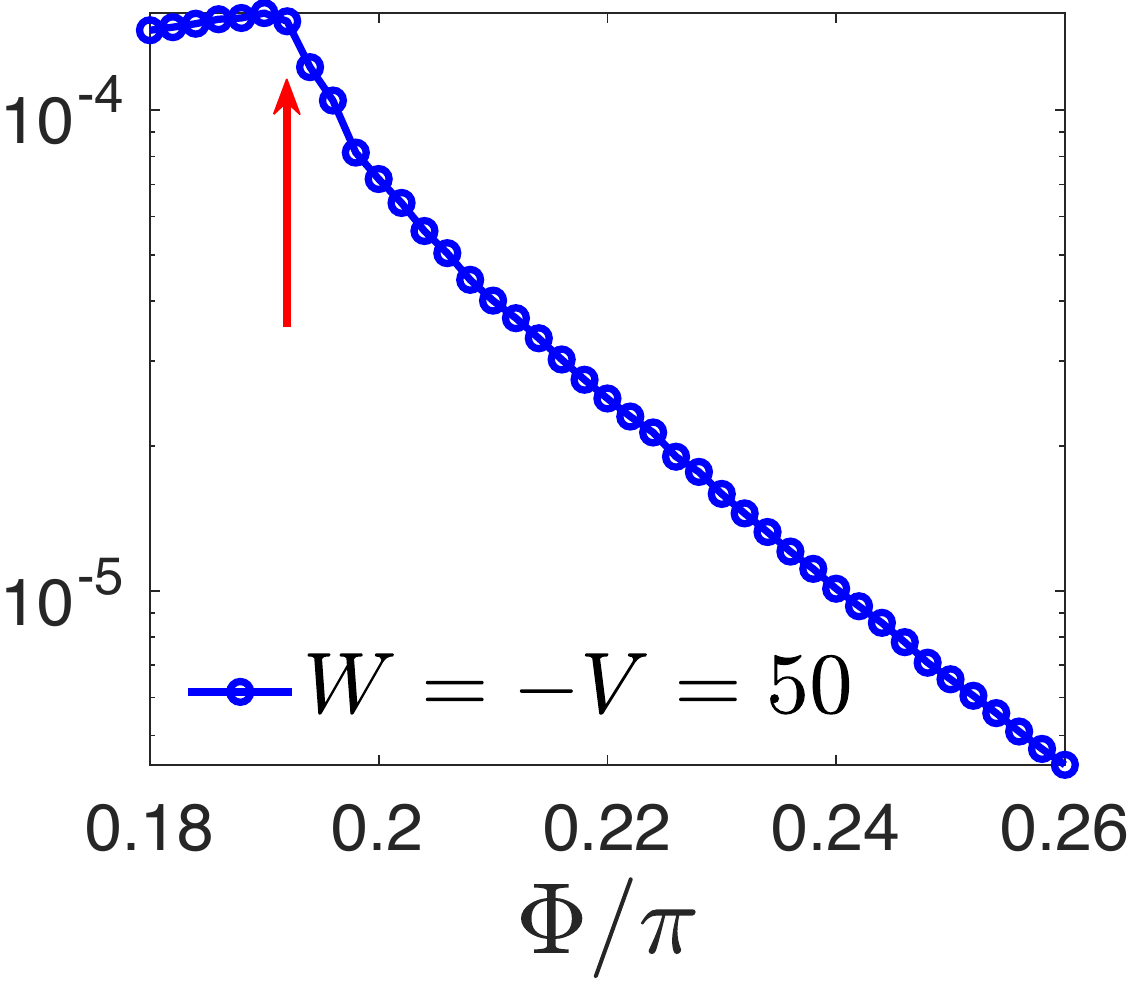}
\caption{Square pair current in log-linear scale, computed as in Eq.~\eqref{eq:squarecoopercurrent}, for the data as in Fig.~\ref{fig:averagecooperdensity2}, panel \textbf{(c)}. Here, the data are (Left) for $W=-V=5\,t$, and (Right) $W=-V=50\,t$. We recall that the data were taken at $L=128$ and $N=32$, using a bond link $M=120$. We see that the current reaches a maximum value around the critical value $\Phi=\Phi_c$, where $\delta n_B$ starts to be nonzero [Fig.~\ref{fig:averagecooperdensity2}, panel \textbf{(c)}], and rapidly decreases for $\Phi>\Phi_c$. The value of $\Phi_c$ indicated by the red arrows is estimated from Fig.~\ref{fig:averagecooperdensity2}, panel \textbf{(c)}.}
\label{fig:coopercurrent}
\end{figure}

\subsection{Varying the particle density}

In our paper, since we discuss the emergence of a gap in the \emph{antisymmetric} sector of the emergent bosonic pairs, what is important is that we choose a value of $n$ such that we are away from any relevant lattice commensurability condition (i.e., $n=1$), which would create a gap also in the \emph{symmetric} sector. For numerical convenience, we choose $n=1/4$. In order to demonstrate that our qualitative results do not rely on this particular choice, we repeated the numerical calculations with $n=1/8$ (using $L=160$ and $N=20$) and show the numerical results in Fig.~\ref{fig:datanormalizeddensity6}. The other numerical parameters are $W=-V=5\,t$, $t_\perp=0.3\,t$, $M=120$ and $S=3$ sweeps. We indeed see that the same phenomenology as in Figs.~\ref{fig:averagecooperdensity2}, \ref{fig:datanormalizeddensity2}, and \ref{fig:datanormalizeddensity4} arises, apart from the fact that the VDW and CDW appear with higher spatial period due to the smaller value of $n$. This result strengthens the conclusion that the physics discussed in the our paper is not a consequence of the specific value of $n$ that we consider.

\subsection{Varying the inter-leg hopping parameter}
In the numerical simulations presented thus far, we use a single fixed value of $t_\perp$, i.e., $t_\perp=0.3\,t$. We now show the numerical data of $\bar r_B$, density and current configuration along the ladder, and density difference $\delta n_B(\Phi)$, for a different value of $t_\perp$, namely $t_\perp=0.5\,t$, in order to further show that the phenology discussed in this paper does not depend on the specific choice of $t_\perp=0.3\,t$.

The additional data for $t_\perp=0.5\,t$ are shown in Figs.~\ref{fig:datanormalizeddensity7} and~\ref{fig:datanormalizeddensity8}. We report the data for $\bar r_B$, for $L=64$, $N=12$ (i.e., $n=1/4$), in order to keep a reasonable numerical complexity, and we scan $V$ from $V=0$ to $|V|=10\,t$, keeping $V=-W$. We also show the ladder density and current configuration, and density difference $\delta n_B(\Phi)$, for $W=-V=7\,t$. Once again, we observe the same phenomenology as in the previous cases discussed in this paper, with a slight shift in the critical value of $\Phi_c$

In light of all these results, we are confident in concluding that the phenomenology discussed in our paper is not a consequence of a fine tuning of the system parameters.

\section{Numerical results for the inter-leg pair current}
\label{sec:numericalresultscurrent}

In this appendix, we provide an additional evidence of the fact that the low-energy physics of our system is indeed given by Eq.~\eqref{eq:pairedfermionshamiltonian}.

An additional observable that we can measure in order to detect the VDW-CDW phase transition is the inter-leg pair current, which on the lattice is defined by the operator
\begin{equation}
\hat J^{(B)}_{\perp,j}=-it_\perp\left(e^{-i\Phi(2j+1)}\,\hat B^\dag_{j,-\frac{1}{2}}\hat B_{j,+\frac{1}{2}}-{\rm H.c.}\right) \,\, .
\end{equation}
Such an operator, at the low-energy level, is sensitive to the pinning of the phase field $\hat\theta_{B,a}$. Therefore, close to the transition, we expect the inter-leg current to be suppressed for $\Phi>\Phi_c$, i.e., when the relative density field $\hat\varphi_{B,a}$ is pinned and therefore $\hat\theta_{B,a}$ becomes strongly fluctuating.

We notice that, by symmetry, the space-average inter-leg current is always zero. The current between the legs can be then quantified for example by the square average current as
\begin{equation}
{\left(\bar J_\perp^{(B)}\right)}^2=\frac{1}{L-2\Delta L}\sum_{j=\Delta L}^{L-\Delta L}{\left|\langle\Psi_{\rm GS}|\hat J^{(B)}_{\perp,j}|\Psi_{\rm GS}\rangle\right|}^2 \,\, ,
\label{eq:squarecoopercurrent}
\end{equation}
where a few sites $\Delta L$ are removed from both chain ends in order to account for open boundary conditions and finite-size effects. We show the numerical data of the square average current in Fig.~\ref{fig:coopercurrent}. The data are computed from the data in Fig.~\ref{fig:averagecooperdensity2}, panel \textbf{(c)}. We see that the current reaches a maximum value around the critical value $\Phi=\Phi_c$, which agrees indeed with the same value of the flux at which $\delta n_B$ starts to be nonzero [see Fig.~\ref{fig:averagecooperdensity2}, panel \textbf{(c)}], and rapidly decreases for $\Phi>\Phi_c$. This result is in agreement with the presence of a relative density order for $\Phi>\Phi_c$, which then implies a strongly fluctuating phase order, detected by the suppression of the current.


%

\end{document}